\RequirePackage{fix-cm}

\documentclass[twocolumn]{svjour3}        

\usepackage{amsfonts,amsmath,amssymb}
\usepackage[margin=1in,footskip=0.5in]{geometry}
\usepackage{booktabs}

\usepackage{scalefnt}

\usepackage[T1]{fontenc}
\usepackage{lmodern}
\usepackage{graphicx}
\graphicspath{ {./Figures/} }
\usepackage{rotating}
\usepackage{color}

\usepackage{siunitx}
\DeclareSIUnit\molar{\mole\per\cubic\deci\metre}
\DeclareSIUnit\Molar{\textsc{m}}

\usepackage{subcaption}		
\usepackage{cancel}
\usepackage{caption}

\newcommand{\pbc}{pre-B\"{o}tC}

\usepackage{natbib}
\usepackage[colorlinks,citecolor=blue,urlcolor=blue,bookmarks=false,hypertexnames=true]{hyperref}

\begin{document}

\title{Dynamics of Ramping Bursts in a Respiratory Neuron Model
}

\author{Muhammad U. Abdulla \and Ryan S. Phillips \and Jonathan E. Rubin
}

\institute{M.U. Abdulla$^{1}$ \at
              	\email{muhammadabdulla@ufl.edu}           
           \and
           R.S. Phillips$^{2,3}$ \at
              	\email{ryanp@pitt.edu}
           \and
           J.E. Rubin$^{2,3}$ \at
              	\email{jonrubin@pitt.edu}
           \and \at
           		$^{1}$	Department of Mathematics, University of Florida, Gainesville, FL, United States \\
           		$^{2}$	Department of Mathematics, University of Pittsburgh, Pittsburgh, PA, United States \\
           		$^{3}$	Center for the Neural Basis of Cognition, Pittsburgh, PA, United States
}

\date{Received: date / Accepted: date}

\maketitle

\begin{abstract}
Intensive computational and theoretical work has led to the development of mutliple mathematical models for bursting in respiratory neurons in the pre-B\"{o}tzinger Complex (pre-B\"{o}tC) of the mammalian brainstem. Nonetheless, these previous models have not captured the preinspiratory ramping aspects of these neurons' activity patterns, in which relatively slow tonic spiking gradually progresses to faster spiking and a full-blown burst, with a corresponding gradual development of an underlying plateau potential. 
In this work, we show that the incorporation of the dynamics of the extracellular potassium ion concentration into an existing model for pre-B\"{o}tC neuron bursting, along with some parameter updates, suffices to induce this ramping behavior. Using fast-slow decomposition, we show that this activity can be considered as a form of parabolic bursting, but with burst termination at a homoclinic bifurcation rather than as a SNIC bifurcation. We also investigate the parameter-dependence of these solutions and show that the proposed model yields a greater dynamic range of burst frequencies, durations, and duty cycles than those produced by other models in the literature.

\keywords{Neuronal dynamics \and Fast-slow dynamics \and Pre-B\"{o}tzinger Complex \and Ion concentration dynamics \and Persistent sodium current}

\end{abstract}

\section*{Declarations}
	\textbf{Funding.} This work was partially supported by NSF awards DMS-1612913 and DMS-1950195 to JER. Additional funding was provided by the University of Florida through the Wentworth Travel Scholarship and the University Scholars Program.\\
	\textbf{Conflicts of Interest.} The authors have no conflicts of interest to disclose.\\
	\textbf{Data Availability.} Not applicable. (See code availability.) \\
	\textbf{Code Availability.} The XPP and MATLAB codes used in this work will be uploaded to ModelDB, where they will be freely available, upon acceptance of this work for publication.

\section{Introduction}
\label{intro}
	Since the original discovery of respiratory activity in neurons within the pre-B\"{o}tzinger Complex (pre-B\"{o}tC) of the mammalian brainstem \citep{smith1991}, many experimental and computational efforts have focused on characterizing the activity patterns of these neurons.
	Experiments have shown that at least under some conditions, individual \pbc\  respiratory neurons can generate temporally clustered action potentials known as bursts.  Moreover, some \pbc\  neuronal bursts have been shown to depend on a persistent sodium current \citep{butera,delnegro2002,del2005sodium,koizumi2008}, while others require a nonspecific cation, or CAN, current \citep{thobybrisson2001,pena2004}, and combinations of these ion flows can produce various distinctive burst patterns including some that may arise under special conditions such as early in development \citep{jasinski,chevalier2016,wang2020} or during sighs \citep{jasinski,toporikova2015,wang2017}.
	
	Functional respiratory rhythms under normoxic conditions consist of three activity phases, commonly known as inspiration, post-inspiration, and late expiration, the latter two of which together comprise expiration.  
	During respiratory rhythms recorded in various experimental preparations, a subpopulation of glutamatergic pre-BötC neurons, sometimes known as type-1 pre-BötC neurons \citep{rekling1998,gray1999} engages in what is known as preinspiratory (pre-I) activity. These neurons remain silent throughout much of post-inspiration and late expiration, but they begin to activate toward the end of the expiration.  Their activity ramps in intensity as expiration gives way to inspiration and culminates in bursting that continues throughout inspiration; indeed, this pre-I activity pattern is thought to play an important role in initiating the expiration-to-inspiration transition.
	While the gradual intensification of pre-I activity likely involves network mechanisms including positive feedback induced by the recruitment of additional neurons, experiments have shown that even individual burst-capable \pbc\  neurons can generate ramping activity patterns, in which tonic spiking  eventually intensifies and transitions to bursting, under pharmacological blockade of glutamatergic neurotransmission \citep{thobybrisson2001,pena2004}.
	
	Despite the significant work done previously to model \pbc\ neuronal activity, current spiking models do not capture the ramping activity observed in individual \pbc\ neurons.  Moreover,  experiments show that the bursting capability of \pbc\ neurons and networks depends on the extracellular ion concentrations to which they are exposed.  Slices of 250-350 $\mu m$ thickness prepared from the \pbc\ are nonrhythmic at physiological [K$^+$]$_{\rm{ext}}$, but some individual \pbc\ neurons do burst in these conditions \citep{delnegro2001,tryba2003stabilization}, especially if depolarized by a tonic input \citep{smith1991}, and pharmacological blockade of GABA$_{\rm{A}}$ and glycinergic inhibition also allows \pbc\ neurons to burst in these conditions \citep{tryba2003stabilization}.  In contrast to these results, however, modeling that explains how different extracellular potassium concentrations can produce corresponding forms of \pbc\ activity has led to the conclusion that, according to existing modeling frameworks, individual \pbc\ neurons should not be able to burst at physiologically relevant extracellular potassium concentrations \citep{bacak}.
	In this paper, we revisit these issues, producing and analyzing what is to our knowledge the first Hodgkin-Huxley (HH) style model for ramping bursts of \pbc\  neurons in the absence of rhythmic drive and inhibitory inputs.  Importantly, our model does not require tuning outside of physiological parameter ranges in order to produce bursting dynamics.
	
	Many of the previous models that inspired this work were also posed in the HH framework, in which a system of nonlinear ordinary differential equations based on Kirchoff's and Ohm's laws represents the temporal evolution of voltage along with a collection of variables modeling the voltage-dependent activation and inactivation levels of transmembrane ionic currents.
	In addition to these variables, HH models include a variety of parameters, representing quantities associated with currents such as time constants, half-activation levels, and reversal potentials.
	Neuronal spikes last just a few milliseconds, whereas inspiratory bursts are much longer events, lasting up to multiple seconds under some experimental conditions. 
	Despite the presence of ionic pumps and glial cells that regulate intra- and extracellular ion concentrations, respectively, spiking that continues over such prolonged periods can lead to significant changes in the ion concentrations that impact neurons \citep{frohlich2008potassium,barreto,kueh2016na+}.
	Given this phenomenon and the knowledge that \pbc\ respiratory neuron activity patterns strongly depend on extracellular potassium concentration, we hypothesized that the dynamics of potassium ions could be central to the emergence of ramping activity in individual \pbc\ neurons.
	The key innovation in our work relative to past \pbc\ neuron models is that we have augmented the HH modeling framework with this ionic dynamics.  In this paper, we show that combining these components yields a neuronal model that successfully produces ramping dynamics.  Applying fast-slow decomposition and associated bifurcation analysis, we explain the mechanisms underlying this activity pattern, which we find represents a form of parabolic bursting. Furthermore, we use direct simulations to explore the robustness and tunability of the bursting dynamics in our model, and we perform additional analysis to elucidate how transitions between bursting and other forms of activity occur as certain model parameters are varied.
	
\section{Model}
	\subsection{Voltage dynamics}
	\label{section:voltagedynamics}
		We consider a model that depicts the spiking behavior of an isolated neuron in the \pbc.
		It is formulated similarly to other HH-style models \citep{huxley} and depends on a persistent sodium current to trigger bursting \citep{butera}. Our model is based heavily on a model presented by \citet{bacak}, augmented with some crucial modifications.
	
		In this model, the membrane potential ($V$) is governed by the current balance equation:
		\begin{equation} \label{eq:capacitance}
			C \cdot \frac{dV}{dt}=-\left( I_{Na}+I_{NaP}+I_{K}+I_{L}+I_{Syn} \right).
		\end{equation}
	
		The membrane currents in (\ref{eq:capacitance}) include: the fast sodium current $I_{Na}$, the persistent sodium current $I_{NaP}$, the delayed rectifier potassium current $I_{K}$, the leakage current $I_{L}$, and the synaptic current $I_{Syn}$.  These membrane currents are drawn from previous work \citet{butera,bacak}, and are represented as follows:
		\begin{align}
			I_{Na} &= \bar{g}_{Na} \cdot (m_{Na})^{3} \cdot h_{Na} \cdot (V-E_{Na}), \label{eq:ina} \\
			I_{NaP} &= \bar{g}_{NaP} \cdot m_{NaP} \cdot h_{NaP} \cdot (V-E_{Na}), \label{eq:inap} \\
			I_{K} &= \bar{g}_{K} \cdot n^{4} \cdot (V-E_{K}), \label{eq:ik} \\
			I_{L} &= \bar{g}_{L} \cdot (V-E_{L}), \label{eq:il} \\
			I_{Syn} &= \bar{g}_{Syn} \cdot (V-E_{Syn}). \label{eq:isyn}
		\end{align}
	
		Note that we model a single neuron, and $I_{Syn}$ is a tonic synaptic current with time-independent conductance,  $\bar{g}_{Syn}$, representing a steady level of drive from other sources, such as brainstem feedback pathways.  This form of synaptic current is appropriate for this study, since we are interested in rhythmicity that can emerge due to intrinsic neuronal dynamics, without contributions from time-varying inputs.
		
	\subsection{Sodium and potassium currents}
		\label{section:na_k_curents}
		The currents $I_{Na}$, $I_{NaP}$, and $I_K$ are given as products of maximal conductances, gating variables, and restoring currents. Each of the sodium gating variables $x \in \{m_{Na}, h_{Na}, m_{NaP}, h_{NaP}\}$ satisfies the equation
		\begin{equation} \label{eq:tau}
			\tau_{x}(V) \cdot \frac{dx}{dt} = x_{\infty}(V) - x,
		\end{equation}
		where
		\begin{align*}
			x_{\infty}(V) &= \left[ 1+\exp \left( (V_x-V) / k_x \right) \right]^{-1}, \\
			\tau_x(V) &= \bar{\tau}_x / \left[ \cosh \left( (V-V_{\tau_x}) / k_{\tau_{x}} \right) \right].
		\end{align*}
			
		The parameter values used for these equations, with corresponding sources and rationales, are all presented in Appendix \ref{constants}.
		
		The potassium current only has activation gates, represented by the variable $n$,  which also satisfies equation (\ref{eq:tau}).
		For $n_{\infty}(V)$ and $\tau_{n}(V)$, we use the formulation
		\[		n_{\infty}(V) = \frac{\kappa_{1}(V)}{\kappa_{1}(V)+\kappa_{2}(V)}, \; \;
		\tau_{n}(V) = \frac{1}{\kappa_{1}(V)+\kappa_{2}(V)},
		\]
		where $\kappa_{1}(V)$ and $\kappa_{2}(V)$ are the following voltage-dependent functions, taken from \citet{bacak,huguenard}:
		\begin{align*}
			\kappa_{1}(V) &= \frac{n_{A} \cdot (n_{A_{V}} + V)}{1 - \exp{\left( -(n_{A_{V}}+V)/n_{A_{k}} \right)}}, \\
			\kappa_{2}(V) &= n_{B} \cdot \exp{\left( -(n_{B_{V}}+V)/n_{B_{k}} \right)}.
		\end{align*}
		The constants $n_{A}$, $n_{B}$, $n_{A_{V}}$, $n_{B_{V}}$, $n_{A_{k}}$, and $n_{B_{k}}$ are discussed in Appendix \ref{constants}.
	
		The reversal potential for potassium ions, denoted $E_{K}$, is viewed as a function of the dynamic variable [ $K^{+} ]_{out}$, and modeled through the Nernst equation approximated at body temperature.
		\begin{equation} \label{eq:nernst}
			E_{K} = 26.7 \cdot \log \frac{ [ K^{+} ]_{out} }{ [ K^{+} ]_{in} },
		\end{equation}
		Note that $[ K^{+} ]_{in}$ is taken to be a constant value. The justification for this approximation is discussed in Sect. \ref{dynamicions}. Internal and external sodium ion concentration, and thus also the sodium reversal potential $E_{Na}$, are taken as constants in this model as in the previous literature \citep{bacak}, with values listed in Appendix \ref{constants}.
		
		\subsection{Ion regulation and dynamics.} \label{dynamicions}
		The crucial difference between our model and the model presented in \citet{bacak} is the inclusion of dynamics in the concentration of extracellular potassium ions, denoted $[K^{+}]_{out}$. 
		
		Experimental data has long indicated that neuronal activity causes fluctuations in $[K^{+}]_{out}$, with maximum increases of roughly $0.8$ \si{\milli\Molar} per spike, which can nearly double the  $[K^{+}]_{out}$ local to a neuron \citep{baylor}.
		
		Experimental manipulations that increase the extracellular potassium concentration are commonly performed in \textit{in vitro} studies to increase neural excitability and induce bursting behavior. A typical approach is to bathe slides containing neuronal tissue in highly concentrated $K^{+}$ solution. 
		The variations of $[K^{+}]_{out}$ due to neural activity and other factors, however, imply that this bath concentration is not equivalent to what we present as the $[K^{+}]_{out}$ variable. 
		Throughout this paper, $[K^{+}]_{out}$ represents the approximate localized concentration of $K^{+}$ in the vicinity of an individual neuron, while $K_{bath}$ represents the concentration of potassium in the bathing solution, toward which $[K^{+}]_{out}$ would naturally evolve over time in the absence of neuronal activity and glial effects.  
		This diffusion of the dynamic $[K^{+}]_{out}$ variable towards $K_{bath}$ is modeled as a molar current of the form discussed in \citet{barreto}:
		\begin{equation} \label{eq:diffusion}
			\tilde{I}_{diff} =  \frac{1}{\tau_{diff}}([K^{+}]_{out} - k_{bath}),
		\end{equation}
		where $\tau_{diff}$ represents the corresponding time constant. To simulate reasonable physiological conditions, $k_{bath}$ was set to $4$ \si{\milli\Molar} \citep{barreto}.
		
		Glial cells also play an active role in decreasing the concentration of $K^{+}$ external to neurons \citep{newman}.  The  effects of the glia on this concentration are also modeled as molar currents in the style of \citet{barreto}, with  maximal rate $\bar{G}$, half-activation potassium concentration $\bar{K}$, and steepness factor $z_{k}$ as follows:
		\begin{equation} \label{eq:glia}
			\tilde{I}_{glia}=\frac{\bar{G}}{1+e^{z_{k} \cdot (\bar{K} - [K^{+}]_{out})}}.
		\end{equation}
		Note that neither diffusion nor glial cells move ions across the neuronal membrane, and thus the currents $\tilde{I}_{diff}, \tilde{I}_{glia}$ do not appear in the voltage equation.
		
		Finally, increases in $[K^{+}]_{out}$ are driven by the action potentials of the neuron. The potassium current $I_{K}$ derives from the movement of potassium ions across the neural membrane. The resulting changes in potassium concentration are therefore proportional to $I_K$.  The proportionality factor is the product of two constants.  One of these terms, $\gamma$, represents the ratio of the  time-derivative of the internal ion concentration to the corresponding membrane current and is derived in Appendix \ref{gamma}.
		The second term, $\beta$, represents the ratio of the internal neuron volume to the localized external volume that determines the reversal potential across the neural membrane. While previous authors have used a value of approximately 7 for this ratio \citep{barreto, somjen}, there is clearly some ambiguity in estimating $\beta$ (as well as $\gamma$) and  our model uses $\beta = 14.555$. Since the parameter $\gamma$ is proportional to changes in internal concentration, changes in external concentration must be proportional to a factor of $\gamma \beta$. Putting these factors together, we model the dynamics of localized external potassium concentration as
		\begin{equation} \label{eq:dk_dt}
			\frac{d[K^{+}]_{out}}{dt} = \gamma \beta I_{K} - \tilde{I}_{diff} - \tilde{I}_{glia}.
		\end{equation}
	
		Also, it is important to note that in this model, $[K^{+}]_{in}$ is approximated as being a constant value, despite the fact that $K^{+}$ ions inside the neuron flow through the neural membrane via the $I_{K}$ current and increase $[K^{+}]_{out}$. 
		The change in external $K^{+}$ concentration, which is under $2$ \si{\milli\Molar} per burst in this model, would only correlate to a decrease of $0.137$ \si{\milli\Molar} in internal $K^{+}$ concentration. This is negligible on the scale of bursting behavior of an individual neuron, as it constitutes only a small fraction of the initial $[K^{+}]_{in}$ value of $150$ \si{\milli\Molar}. This approximation was also used in a previous neuronal bursting model with dynamic ion concentrations, based on the argument that changes in $[K^{+}]_{in}$ are more strongly correlated to fluctuations in internal sodium ion concentration rather than to changes in $[K^{+}]_{out}$ \citep{barreto}.
		
	\subsection{The full model}
		In summary, we arrive at a 7-dimensional model of a neuron, which depicts bursting behavior by connecting the dynamics of membrane potential, sodium and potassium gating and reversal potentials, and ion concentrations. The formulations of these dynamics are based on a combination of previous models of bursting behavior \citep{butera,bacak,barreto}. The differential equations in this system are equations (\ref{eq:capacitance}), (\ref{eq:tau}), and (\ref{eq:dk_dt}); note that in fact we have 5 equations of the form (\ref{eq:tau}), one for each of $m_{Na}$, $h_{Na}$, $m_{NaP}$, $h_{NaP}$, and $n$.
		
\section{Periodic Behaviors in the Model} \label{section:periodic}
	\subsection{Activity patterns} \label{burstdesc}
\begin{figure*}[htbp] 
				\begin{minipage}[b]{0.5\linewidth}
					\centering
					\includegraphics[width=1\linewidth]{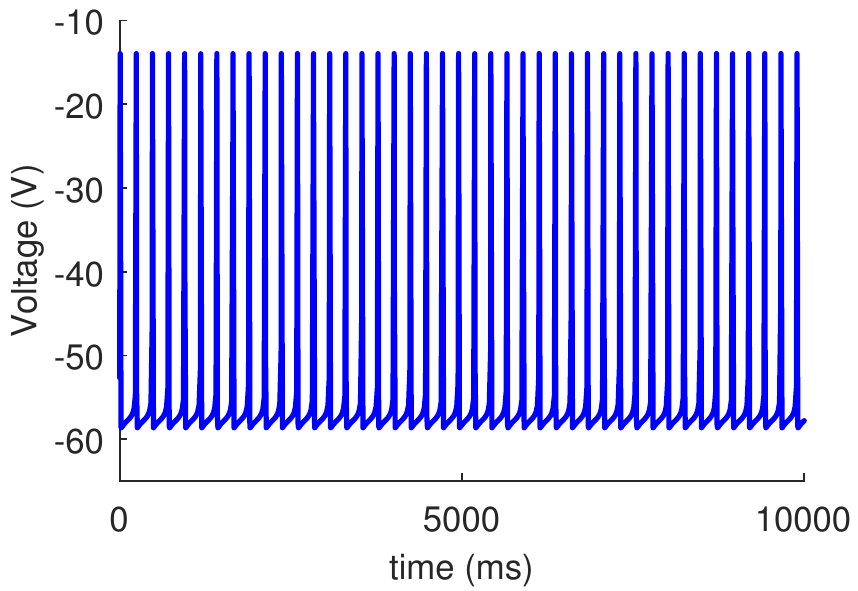}
					\caption*{({\bf A}): $[K^{+}]_{out} = 4.0$ \si{\milli\Molar}, $E_{K} = -96.8$ \si{\milli\volt}}
				\end{minipage}
				\begin{minipage}[b]{0.5\linewidth}
					\centering
					\includegraphics[width=1\linewidth]{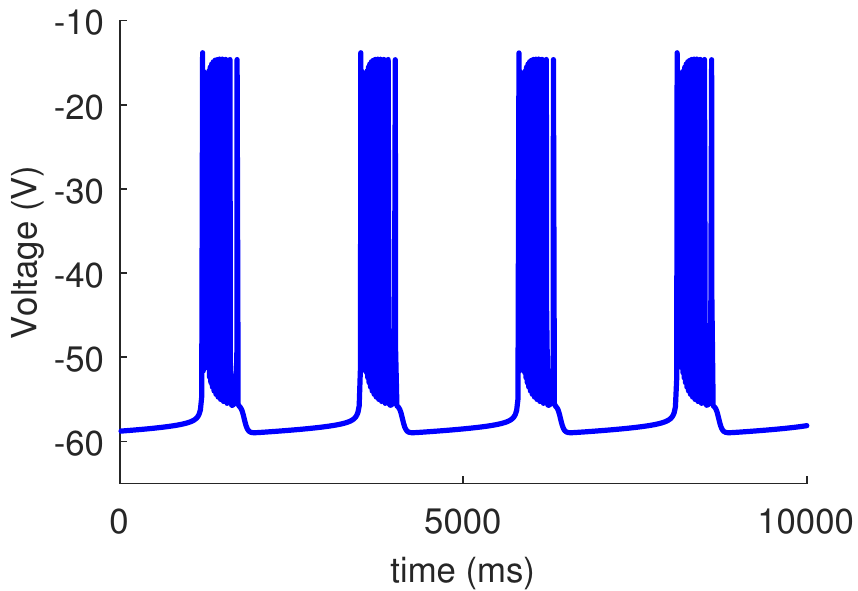}
					\caption*{({\bf B}): $[K^{+}]_{out} = 6.0$ \si{\milli\Molar}, $E_{K} = -85.9$ \si{\milli\volt}}
				\end{minipage}
				\begin{minipage}[b]{0.5\linewidth}
					\centering
					\includegraphics[width=1\linewidth]{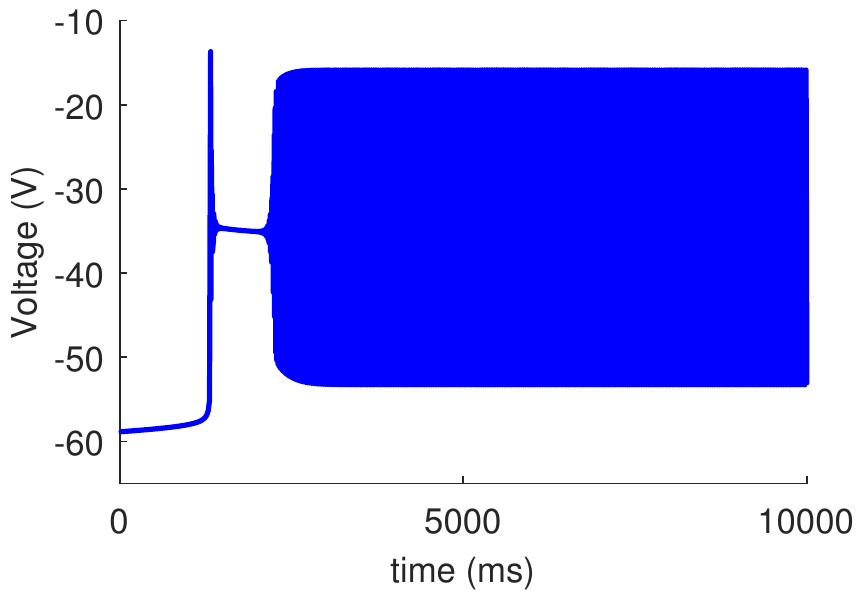}
					\caption*{({\bf C}): $[K^{+}]_{out} = 8.0$ \si{\milli\Molar}, $E_{K} = -78.3$ \si{\milli\volt}}
				\end{minipage}
				\begin{minipage}[b]{0.5\linewidth}
					\centering
					\includegraphics[width=1\linewidth]{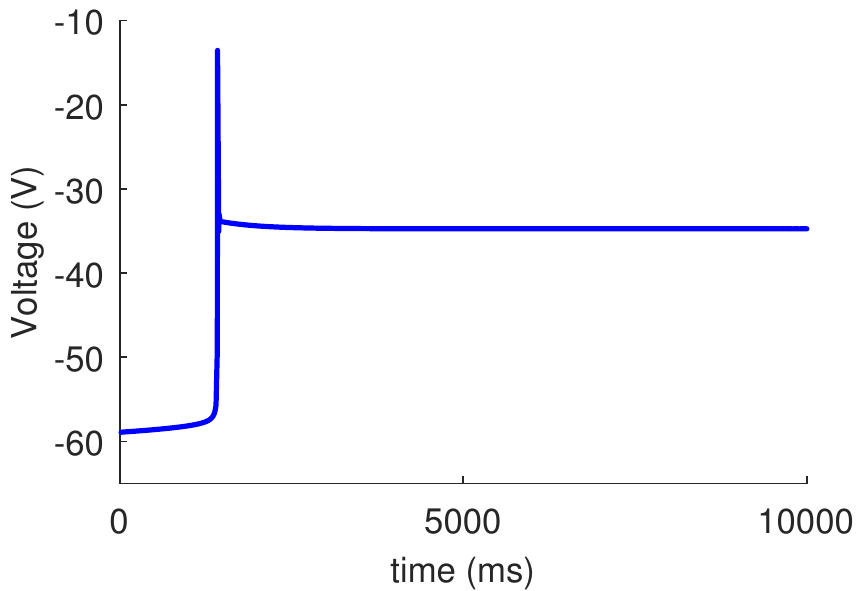}
					\caption*{({\bf D}): $[K^{+}]_{out} = 10.0$ \si{\milli\Molar}, $E_{K} = -72.3$ \si{\milli\volt}}
				\end{minipage}
				\caption{\small{Model \pbc\ neuron activity depends on the potassium reversal potential, $E_K$, which relates to the local external potassium concentration via equation (\ref{eq:nernst}). \textbf{(A)}  At $E_{K} = -96.8$ \si{\milli\volt}, the reversal potential is too low to support bursting, and the neuron remains in a tonic spiking state, characterized by rhythmic spiking at a fixed frequency. \textbf{(B)} At $E_{K} = -85.9$ \si{\milli\volt}, the neuron has surpassed the threshold $E_{K}$ value for bursting and exhibits periods of quiescence alternating with periods of high-frequency spiking riding a depolarized voltage plateau. \textbf{(C)} At $E_{K}= -78.3$ \si{\milli\volt}, after release from a resting level potential, the neuron spikes but cannot fully repolarize and return to a resting state. As a result, it once again enters a tonic spiking state, but with reduced repolarization and a higher frequency relative to (A). \textbf{(D)} Finally, at $E_{K} = -72.3$ $\si{\milli\volt}$, the neuron enters depolarization block with an elevated membrane potential and an absence of spike generation.}}
				\label{behavior}
			\end{figure*}

			To match experimental data, a \pbc\ neuron model must demonstrate a range of activity patterns across different conditions.  Previous modeling work showed how different neuronal behaviors occur at different fixed values of the external potassium concentration, and our model reproduces this result in Fig. \ref{behavior}. This agreement is not surprising:  When internal and external $K^{+}$ concentrations are fixed, our model is extremely similar to the model presented in \citet{bacak}, differing only in the values of a few model parameter values, which affect quantitative but not qualitative aspects of the dynamics in this frozen-potassium setting.
			
			Previous work has noted that fixing  $[K^{+}]_{out}$, which is directly related to $E_{K}$ by (\ref{eq:nernst}), at  values sufficiently elevated above physiological levels is enough to induce bursting in a \pbc\ neuron model lacking ion concentration dynamics \citep{bacak}.  Furthermore, modeling of other brain areas revealed a wide array of bursting behaviors when $K^{+}$ and $Na^{+}$ concentrations were allowed to vary dynamically \citep{barreto,erhardt2020}. In this work, we  combine the insights offered by these earlier investigations to model \pbc\ dynamics featuring ramping activity culminating in a burst without imposed elevation of extracellular potassium concentration.
			
\begin{figure*}[htbp] 
				\begin{minipage}[b]{0.5\linewidth}
					\centering
					\includegraphics[width=1\linewidth]{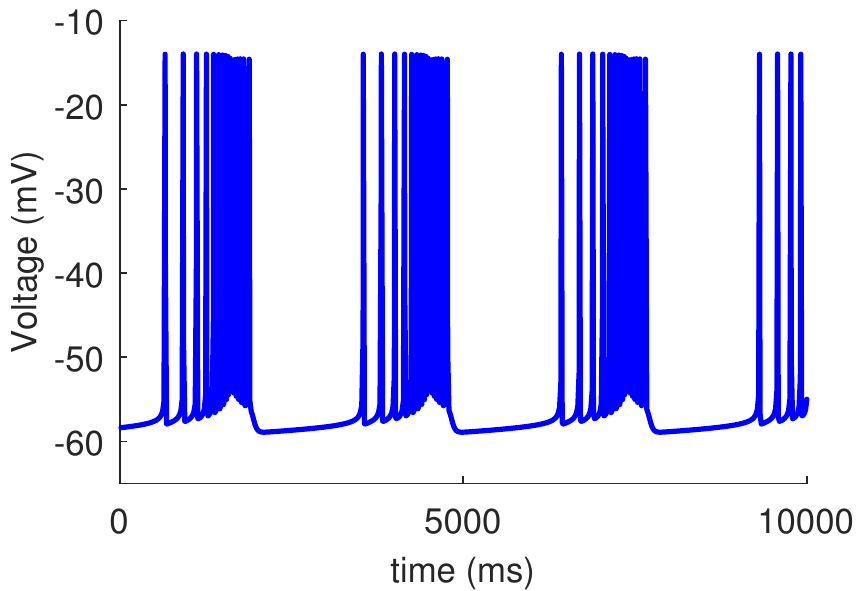} 
					\caption*{({\bf A}): $V$ vs. $t$}
				\end{minipage}
				\begin{minipage}[b]{0.5\linewidth}
					\centering
					\includegraphics[width=1\linewidth]{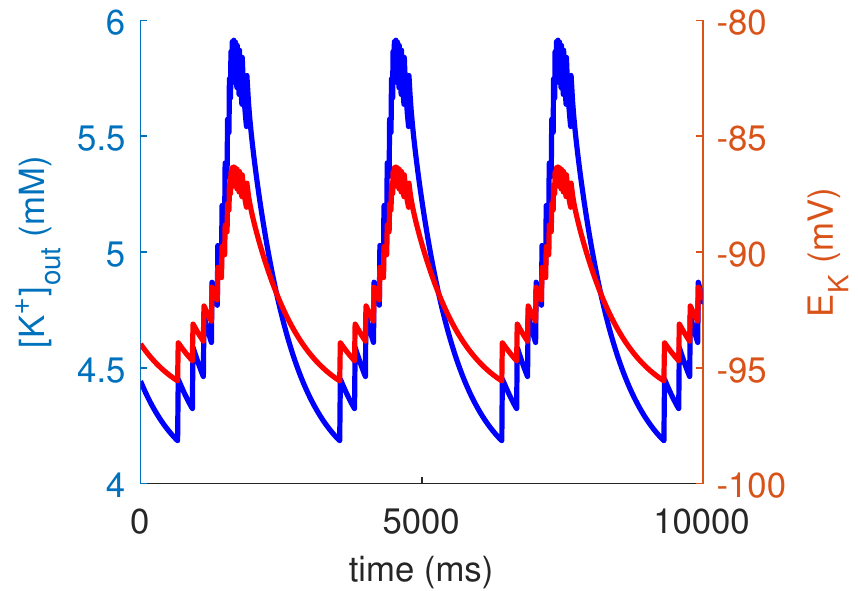}
					\caption*{({\bf B}): $[K^{+}]_{out}$ \& $E_{K}$ vs. t}
				\end{minipage}
				\begin{minipage}[b]{0.5\linewidth}
					\centering
					\includegraphics[width=1\linewidth]{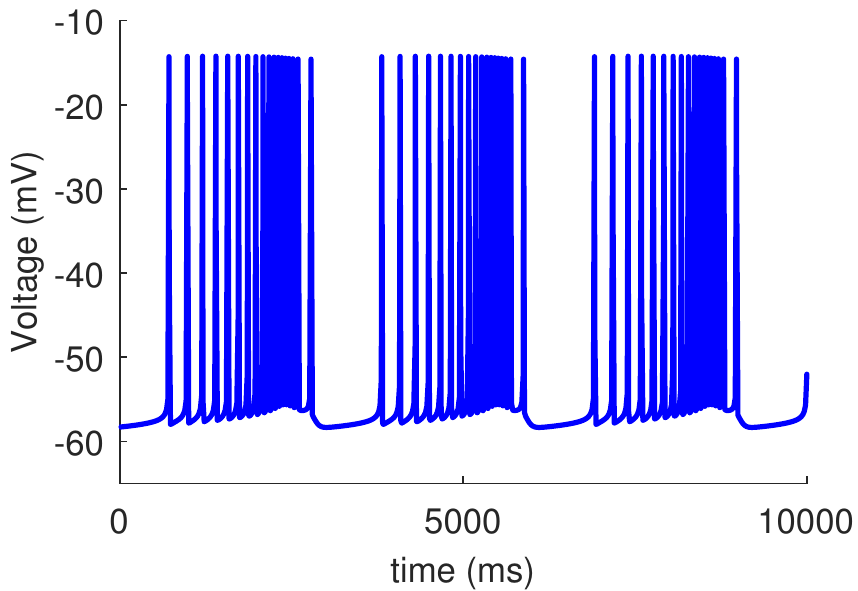}
					\caption*{({\bf C}): $V$ vs. $t$}
				\end{minipage}
				\begin{minipage}[b]{0.5\linewidth}
					\centering
					\includegraphics[width=1\linewidth]{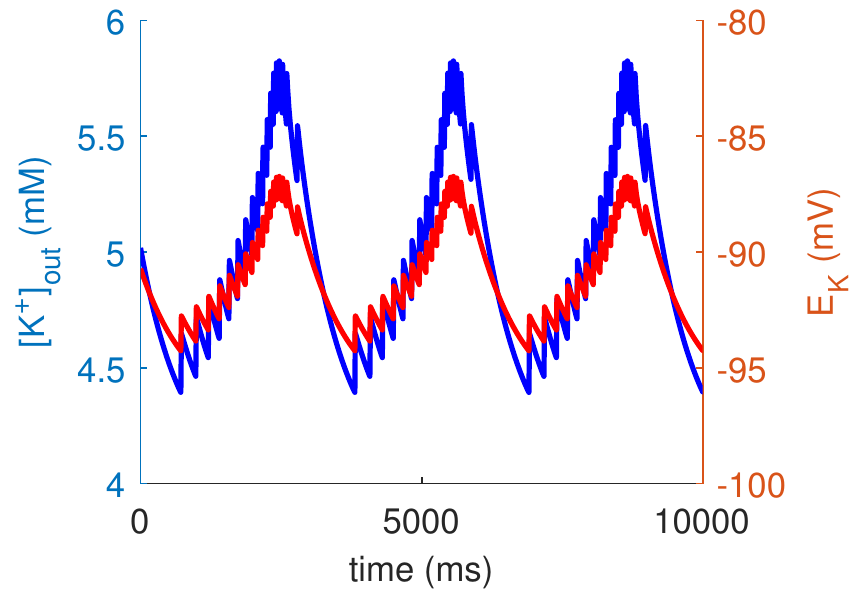}
					\caption*{({\bf D}): $[K^{+}]_{out}$ \& $E_{K}$ vs. t}
				\end{minipage}
				\caption{\small{Typical bursting trajectories of our preinspiratory \pbc\ neuron model with $[K^{+}]_{out}$ dynamics. \textbf{(A)} The membrane potential (in $\si{\milli\volt}$) plotted over time (in $\si{\milli\second}$). Note that over the course of each burst, the frequency of the spiking tends to increase and then decrease, while spike amplitude has the opposite trend (parameter set: $g_{NaP} = 5.0$ \si{\nano\siemens}, $g_{L} = 2.50$ \si{\nano\siemens}, $g_{syn} = 0.365$ \si{\nano\siemens}). \textbf{(B)} The time course of the potassium reversal potential ($E_K$, red) along with external potassium concentration ($[K^+]_{out}$, blue). Note that increases in $E_{K}$ align with increases in spiking frequency. \textbf{(C)} Voltage time course for bursting with reduced conductances (parameter set: $g_{NaP} = 4.5$ \si{\nano\siemens}, $g_{L} = 2.40$ \si{\nano\siemens}, $g_{syn} = 0.360$ \si{\nano\siemens}). Note that the spiking frequency increases more gradually over the course of the burst. \textbf{(D)} Time courses of $E_K$ and $[K^+]_{out}$ for this alternative burst waveform.}}
				\label{bursting}
			\end{figure*}

			Indeed, with dynamic extracellular potassium levels, our model produces distinctive ramping bursts as shown in Fig. \ref{bursting}, matching a pattern seen experimentally in \pbc\ neurons; the slow spiking on a gradually increasing voltage plateau at the start of each burst active phase is referred to in the literature as ``preinspiratory activity''.  These bursts include periods of quiescence, during which $[K^{+}]_{out}$ remains on the low end of physiologically observed levels, corresponding to low values of $E_K$, by equation (\ref{eq:nernst}). Numerical simulations show that $E_K$ slowly increases during this phase until spiking emerges.
			As in other HH-type models, each spike involves dynamics of the sodium and potassium currents, $I_{Na}$ and $I_{K}$, respectively. The ion flows associated with these currents gradually increase $[K^{+}]_{out}$. Although glia and diffusion regulate external $K^{+}$ concentrations, the strengths of these repolarization currents depend on $[K^{+}]_{out}$ as depicted in equation (\ref{eq:dk_dt}). At low concentrations, the glia are almost inactive and diffusion is too weak to bring $[K^{+}]_{out}$ back to equilibrium. A positive feedback loop results, such that as the neuron continues to spike and $[K^{+}]_{out}$ continues to increase substantially above baseline values.
			This rise in $[K^{+}]_{out}$ is cut off by the nonlinear rise in the strength of diffusion and glial currents as in equations (\ref{eq:diffusion})-(\ref{eq:glia}). The overall increase of $E_{K}$ is enough to trigger bursting behavior in the neural cell, however, and this bursting continues until some time after $[K^{+}]_{out}$ saturates. Furthermore, as demonstrated experimentally \citep{delnegro2001} and discussed below (cf.  Fig. \ref{lowglfixedek}), the spiking frequency increases with $[K^{+}]_{out}$. Thus, the increasing $[K^{+}]_{out}$ during the build-up of a burst also provides a mechanism for a \textit{ramping effect}, where the spiking frequency gradually increases from an initial slow tonic spiking until a burst is established.
			The exact geometry of the burst pattern depends on various parameters, including conductance strengths.  For example, with a reduction in $g_{NaP}, g_L$, and $g_{syn}$ the bursting pattern changes to feature a more gradual increase in spike frequency and a less pronounced drop in spike amplitude during the burst (Fig. \ref{bursting}, bottom).
			
		\subsection{Fast-slow decomposition analysis}
			Neuronal bursting results from dynamics occurring across two or more distinct timescales. Voltage spikes occur on a fast timescale. Transitions between the spiking state and quiescent state within the bursting regime, as well as the gradual oscillation of $[K^{+}]_{out}$ over the course of a burst, depend on slow timescale dynamics. In our model, a positive feedback loop between the slow subsystem and the fast subsystem causes a buildup in external $K^{+}$ concentration and a gradual increase in spike frequency during the active phase of a burst. The variation in $E_{K}$ values that results affects the timing of the transition from the active spiking state to the quiescent state within each burst.
			
			A fast-slow decomposition is a standard mathematical approach to elucidate the details of multiple timescale dynamics in bursting \citep{bertramrubin}. We begin a fast-slow decomposition by noting that $h_{NaP}$ and $[K^+]_{out}$ evolve significantly more slowly that the other variables in the model.
			Hence, the full model can be considered as having 4 fast variables, comprising a fast subsystem, and 2 slow variables, constituting a slow subsystem.
			
			A standard approach when a model features multiple slow variables, which we follow, is to pick one of these as a primary bifurcation parameter and compute bifurcation diagrams for the fast subsystem with respect to this parameter, while the other slow variables are held frozen at some fixed values.
			This process can then be repeated for various values of  these other slow variables, which are typically selected based on the paths they follow when the full system evolves.
			This approach does not capture certain transitional solution patterns that involve subtle interactions of multiple slow variables or mixing of time scales \citep{vo,teka,wang2016,wang2017,bertramrubin,wang2020}, but it can be an effective way to explain many activity patterns in fast-slow systems nonetheless.
	
			Previous analysis of respiratory neuron models with fixed $E_K$ showed the utility of $h_{NaP}$ as a bifurcation parameter \citep{butera,bacak}, so we make $h_{NaP}$ our initial primary bifurcation parameter as well, and we use XPPAUT \citep{xpp} to consider how the dynamics of the fast subsystem varies with $h_{NaP}$. We repeat this analysis for several values of $[K^+]_{out}$ (and hence of $E_K$). Note that we refer to the fast subsystem together with $h_{NaP}$ as the {\it neuronal system}.
			
			\begin{figure*}[htbp] 
				\begin{minipage}[b]{0.5\linewidth}
					\centering
					\includegraphics[width=1\linewidth]{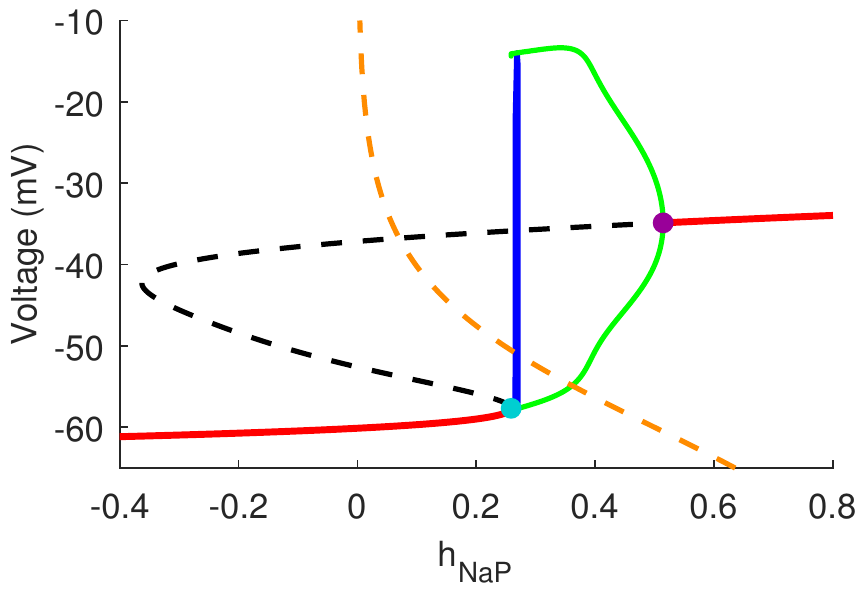}
					\caption*{({\bf A}): $[K^{+}]_{out} = 4.5$ \si{\milli\Molar}, $E_{K} = -93.6$ \si{\milli\volt}.}
				\end{minipage}
				\begin{minipage}[b]{0.5\linewidth}
					\centering
					\includegraphics[width=1\linewidth]{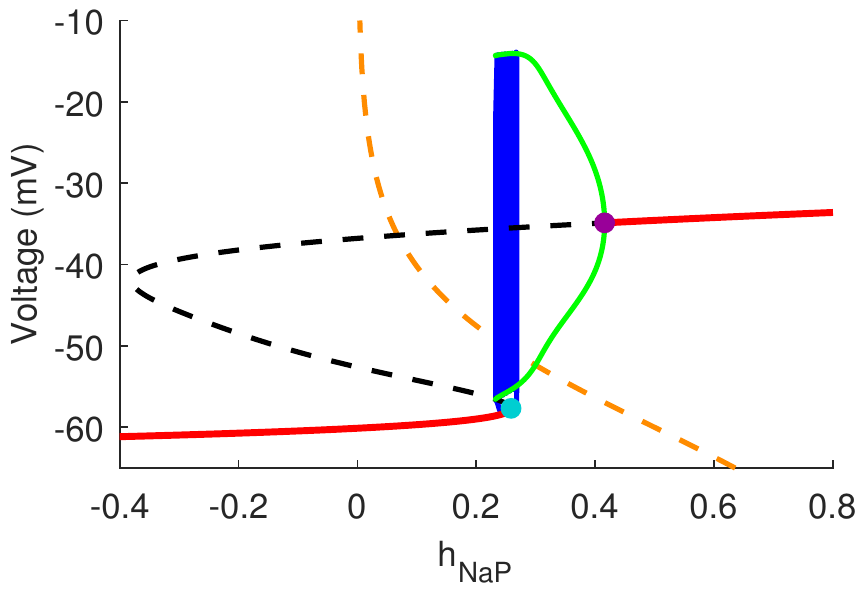}
					\caption*{({\bf B}): $[K^{+}]_{out} = 5.3$ \si{\milli\Molar}, $E_{K} = -89.3$ \si{\milli\volt}.}
				\end{minipage}
				\begin{minipage}[b]{0.5\linewidth}
					\centering
					\includegraphics[width=1\linewidth]{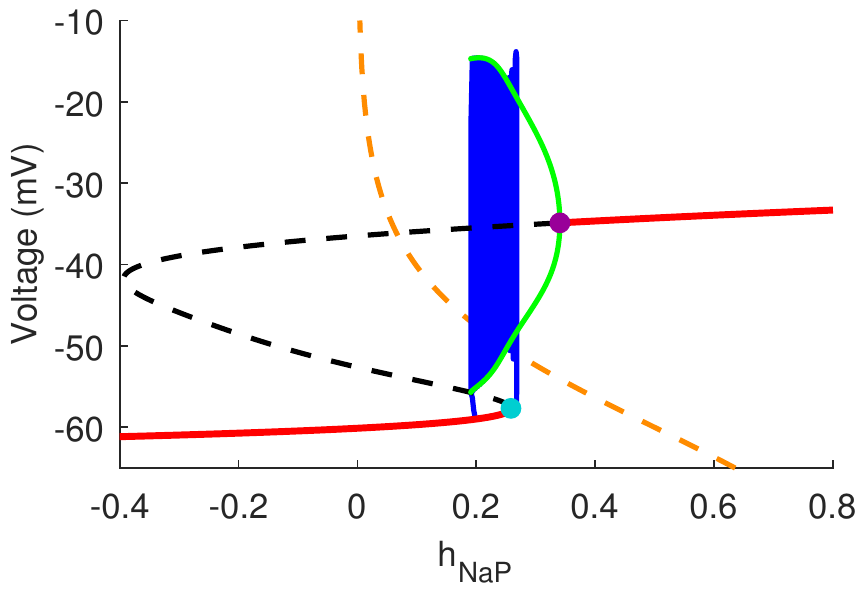}
					\caption*{({\bf C}): $[K^{+}]_{out} = 6.0$ \si{\milli\Molar}, $E_{K} = -85.9$ \si{\milli\volt}.}
				\end{minipage}
				\begin{minipage}[b]{0.5\linewidth}
					\centering
					\includegraphics[width=1\linewidth]{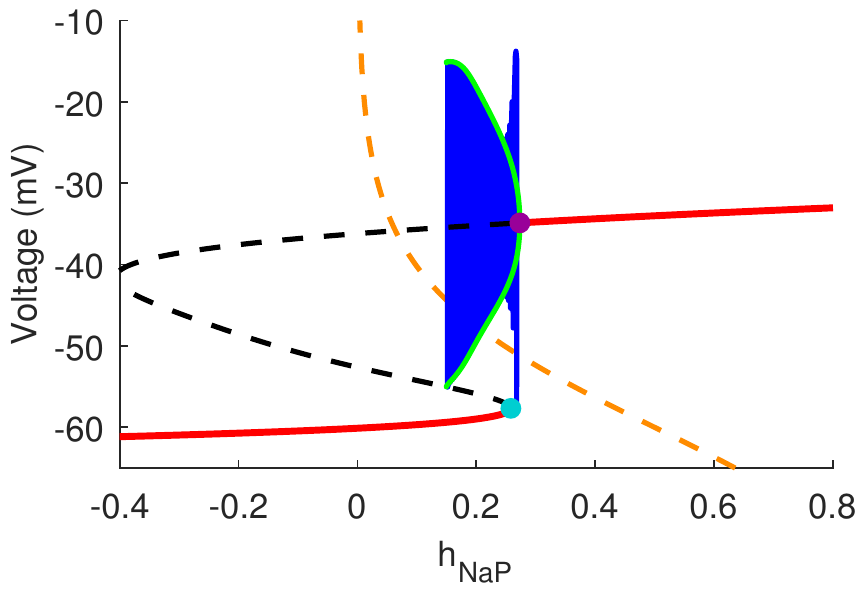} 
					\caption*{({\bf D}): $[K^{+}]_{out} = 6.7$ \si{\milli\Molar}, $E_{K} = -83.0$ \si{\milli\volt}.}
				\end{minipage}
				\caption{\small{Both spiking and bursting states can be realized with fixed $K^{+}$ concentration (cf. \cite{bacak}). In each panel, the solid red curve indicates stable (upper and lower) segments of the critical manifold, while the black dotted curves show unstable segments. The solid purple circle labels an Andronov-Hopf (AH) bifurcation point $h_{NaP}^{AH}$ and the green curve represents the periodic orbit family $\mathcal{P}$ originating from this AH point. The blue curve represents an orbit of the neuronal system starting from a jump up to the active phase. The dashed orange curve is the $h_{NaP}$ nullcline.  \textbf{(A)} At low $E_{K}$, the trajectory converges to a spiking oscillation near the end of $\mathcal{P}$, which occurs in a SNIC bifurcation (light blue) at the lower fold, or knee, of the critical manifold $\mathcal{S}$, with $h=h_{NaP}^{LK}$. \textbf{(B)} As $E_{K}$ increases, $h_{NaP}^{AH}$ decreases and the termination of $\mathcal{P}$ switches to a homoclinic bifurcation at $h_{NaP}^{HC}$.  The neuronal system switches from spiking to bursting. \textbf{(C)} As $E_{K}$ continues to increase, $h_{NaP}^{AH}, h_{NaP}^{HC}$ both decrease and the neuronal system's bursting trajectory reaches lower $h_{NaP}$ values. The shape of the bursting waveform is determined by the relative positions of $h_{NaP}^{LK}$ and $h_{NaP}^{AH}$. \textbf{(D)} For large enough $E_{K}$, $h_{NaP}^{AH}<h_{NaP}^{LK}$; moreover, the relative positions of $\mathcal{P}$ and the $h_{NaP}$-nullcline result in tonic spiking (note the absence of a jump down in the orbit from $\mathcal{P}$ to the lower stable branch of $\mathcal{S}$; also see main text).}}
				\label{2dphase}
			\end{figure*}
		
			Let us start with the parameter set corresponding to Fig. \ref{bursting}A.
			Consider first $[K^{+}]_{out} = 4.5$ \si{\milli\Molar} (Fig. \ref{2dphase}, upper left).  The fast subsystem bifurcation diagram with respect to $h_{NaP}$ includes an S-shaped curve of equilibria, known as the critical manifold $\mathcal{S}$, including two stable segments (red solid), one a hyperpolarized branch corresponding to quiescence and the other a depolarized segment corresponding to depolarization block.  The lower stable branch ends in a saddle-node bifurcation that we call the lower knee of $\mathcal{S}$, with  $h=h_{NaP}^{LK}$, while the upper segment destabilizes at even larger $h_{NaP}$ at a supercritical Andronov-Hopf (AH)  bifurcation, with $h=h_{NaP}^{AH}$.  These bifurcation values do depend on $[K^{+}]_{out}$, but we suppress this dependence in our notation.  The family of stable periodic orbits, $\mathcal{P}$, born in the AH bifurcation continues for decreasing $h_{NaP}$ until terminating in a SNIC bifurcation at the lower knee.  When the neuronal system, consisting of the fast subsystem along with the slow $h_{NaP}$ dynamics, is simulated with $[K^{+}]_{out}$, and thus $E_{K}$, still frozen, the system exhibits periodic tonic spiking in which $h_{NaP}$ hovers near a particular value and the voltage of the cell oscillates along the associated part of the periodic orbit family in the bifurcation diagram. Past work has shown that this tonic spiking results when the weak leftward drift in $h_{NaP}$ during the part of each oscillation when the trajectory lies above the $h_{NaP}$-nullcline (dashed orange) in $(h_{NaP},V)$-space exactly balances the weak rightward drift when the trajectory is below the $h_{NaP}$-nullcline \citep{bacak}.
			
			When $[K^{+}]_{out}$ is fixed at the larger value of  $5.3$ \si{\milli\Molar}, the fast subsystem bifurcation diagram remains similar but the termination of the periodic orbit family decouples from the saddle-node bifurcation; that is, the termination now occurs at a homoclinic bifurcation, with $h=h_{NaP}^{HC}$, instead of at a SNIC. The shift in the periodic orbit family due to the selection of a new $E_K$ value also changes its relation to the position of the $h_{NaP}$-nullcline and its shape.  As a result, the trajectory of the neuronal system drifts in the direction of lower $h_{NaP}$ as spiking occurs until it reaches the $h_{NaP}$ value of the homoclinic bifurcation and returns to the silent, non-spiking phase.  Thus, this system  produces square-wave bursting, also known as fold-homoclinic bursting \citep{izhikevich} (Fig. \ref{bursting}B).
			
			As we consider progressively larger (less hyperpolarized) values of $[K^{+}]_{out}$, $\mathcal{P}$, $h_{NaP}^{AH}$, and $h_{NaP}^{HC}$ all move to smaller $h_{NaP}$ values.
			Moreover, the curve of maximal voltages along the periodic orbit family continues to change shape, becoming monotone decreasing in $h_{NaP}$ instead of non-monotonic as previously.
			When $[K^{+}]_{out} = 6.0$ \si{\milli\Molar}, for example, the neuronal system continues to produce bursting dynamics, but with bursts of longer duration and more spikes per burst than previously (Fig. \ref{bursting}C). As $h_{NaP}^{AH}$ becomes closer to $h_{NaP}^{SN}$, the initial spikes within each burst have a large amplitude but subsequent spikes are smaller, as the orbit converges down to small-amplitude fast subsystem periodic orbits near the AH point; as time continues to evolve, spikes become larger again, as the bursting orbit travels toward the homoclinic, where the fast subsystem periodics have larger amplitude.  This decreasing-increasing trend in spike amplitudes becomes more pronounced as $[K^{+}]_{out}$ increases and $h_{NaP}^{AH}$ moves to successively smaller $h_{NaP}$.
			
			Finally, at a $[K^{+}]_{out}$ value above a certain threshold, the neuronal system no longer produces bursting behavior. For example, for $[K^{+}]_{out}=6.7$ \si{\milli\Molar}, the AH point now lies to the left of the saddle-node point. Hence, if we start a trajectory in the silent phase, then after $h_{NaP}$ grows and reaches the SN point to initiate spiking, the initial decline in spike amplitude is particularly pronounced, as the trajectory initially converges toward the depolarized branch of fast subsystem equilibria (Fig. \ref{bursting}D).  Furthermore, thanks to the more extreme leftward position of the periodic orbit family,  the spikes that occur at low $h_{NaP}$ spend significant time below the $h_{NaP}$-nullcline in the $(h_{NaP},V)$ plane, allowing the corresponding rightward drift in $h_{NaP}$ to balance the leftward drift that occurs when voltage is more depolarized.   Thus, the trajectory becomes pinned and oscillates along a particular  fast subsystem periodic orbit indefinitely, as it did for $[K^{+}]_{out} = 4$ \si{\milli\Molar}, and the neuron remains in a tonic spiking state.
			
\begin{figure*}[htbp] 
				\begin{minipage}[b]{0.5\linewidth}
					\centering
					\includegraphics[width=1\linewidth]{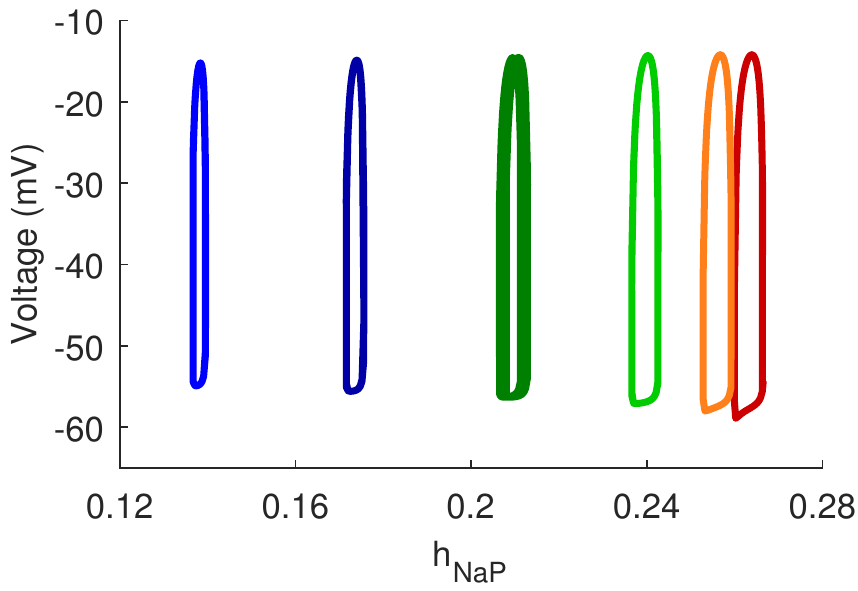} 
					\caption*{({\bf A})} 
				\end{minipage}
				\begin{minipage}[b]{0.5\linewidth}
					\centering
					\includegraphics[width=1\linewidth]{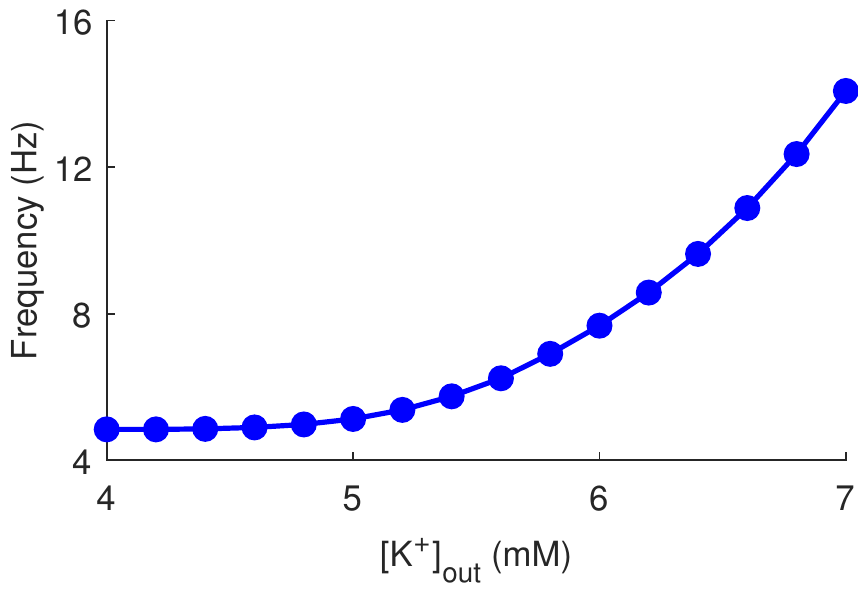}
					\caption*{({\bf B})} 
				\end{minipage}
				\caption{\small{Limit cycles of the \pbc\ neuron model with fixed $[K^{+}]_{out}$ for the parameter set depicted in Fig. \ref{bursting}. \textbf{(A)} Starting from the rightmost loop and moving leftwards, $[K^{+}]_{out}$ = $4.0, 4.6, 5.2, 5.8, 6.4, 7.0$ \si{\milli\Molar}.  Note that at $[K^{+}]_{out} = 5.0$ \si{\milli\Molar}, the trajectory of the neuron appears thicker than the other trajectories. Here, the neuron came close to undergoing a period doubling bifurcation and establishing a distinctive $2$-spike bursting pattern. Nonetheless, the neuron was unable to achieve bursting with robustness anywhere near that of the dynamic $[K^{+}]_{out}$ model. \textbf{(B)} The spiking frequency of the periodic orbit of the neuronal dynamics is plotted against the fixed  $[K^{+}]_{out}$ value at which it occurs.}}
				\label{lowglfixedek}
			\end{figure*}

			Next, consider the parameter set with $g_{syn} = 0.360$ \si{\nano\siemens}, $g_{NaP} = 4.5$ \si{\nano\siemens}, and $g_{L} = 2.4$ \si{\nano\siemens}. This reduction in $g_{L}$ is analogous to increasing the excitability of the neuron, as discussed in more detail in Sect. \ref{section:robustness}. In Fig. \ref{bursting}C, the bursting waveform resulting from this parameter set is demonstrated. However, if $[K^{+}]_{out}$ is set to be constant, bifurcation analysis with respect to $h_{NaP}$ shows that the neuronal system is unable to achieve a bursting state at any fixed ion concentration. Fig. \ref{lowglfixedek} displays example trajectories of tonic spiking solutions of the  neuronal dynamics that arise with these parameter values when $[K^{+}]_{out}$ is fixed, which appear as thin closed loops when projected to $\left( V, h_{NaP} \right)$ phase space. 
			
			The gradual rise of $[K^{+}]_{out}$ essentially drags the trajectory of the neuron through the steady states shown in Fig. \ref{lowglfixedek}A in the direction of lower $h_{NaP}$. As these states correspond to higher spiking frequencies, as depicted in Fig. \ref{lowglfixedek}B, the dynamics of $[K^{+}]_{out}$ provides a mechanism for an active phase geometry that features a gradual increase in spiking frequency. Thus, we have shown that even a neuron that can never burst on its own with fixed $[K^{+}]_{out}$ can nonetheless become intrinsically bursting when $[K^{+}]_{out}$ dynamics are taken into account.
			
		\subsection{A closer look at transitions in behavior as $E_{K}$ is varied}
			As illustrated in Fig. \ref{behavior}, if the $K^{+}$ concentration is held fixed, then shifting the $E_{K}$ value has a clear effect on the long-term periodic behavior of the model neuron. Each periodic behavior, whether tonic spiking or bursting, can be depicted as a stable limit cycle projected to the $(V, h_{NaP})$ phase space. As shown in Fig. \ref{2dphase}, with increases in $E_K$, the stable oscillation switches from tonic spiking to bursting, and then, with additional increases, from bursting back to spiking.  Which behavior arises depends on whether the periodic orbit family of the fast subsystem terminates in a SNIC bifurcation or a homoclinic bifurcation and on where this termination lies relative to the $h_{NaP}$ nullcline.
			
\begin{figure*}[htbp] 
				\begin{minipage}[b]{1\linewidth}
					\centering
					\includegraphics[width=1\linewidth]{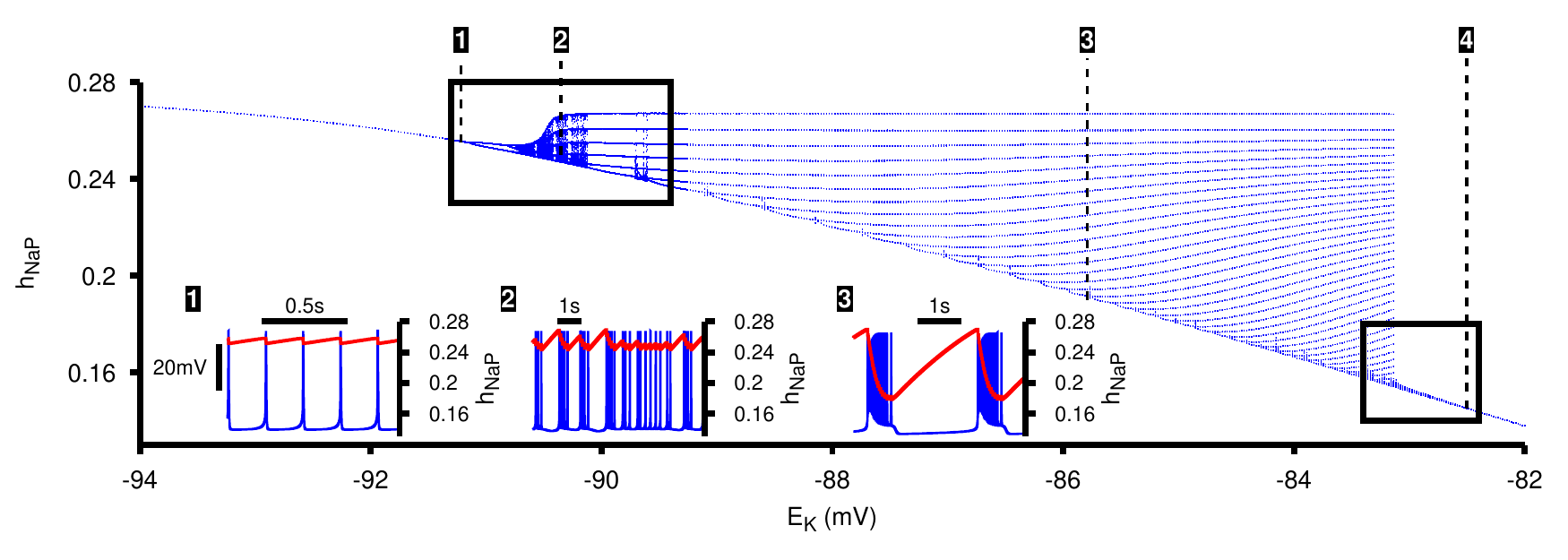}
					\caption*{({\bf A})}
				\end{minipage}
				
				\begin{minipage}[b]{0.5\linewidth}
					\centering
					\includegraphics[width=1\linewidth]{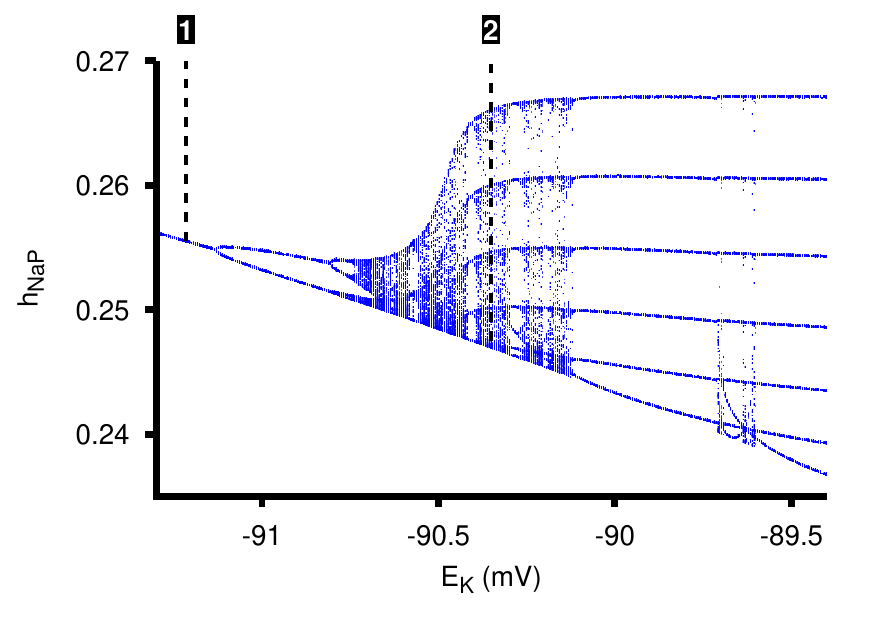}
					\caption*{({\bf B})}
				\end{minipage}
				\begin{minipage}[b]{0.5\linewidth}
					\centering
					\includegraphics[width=1\linewidth]{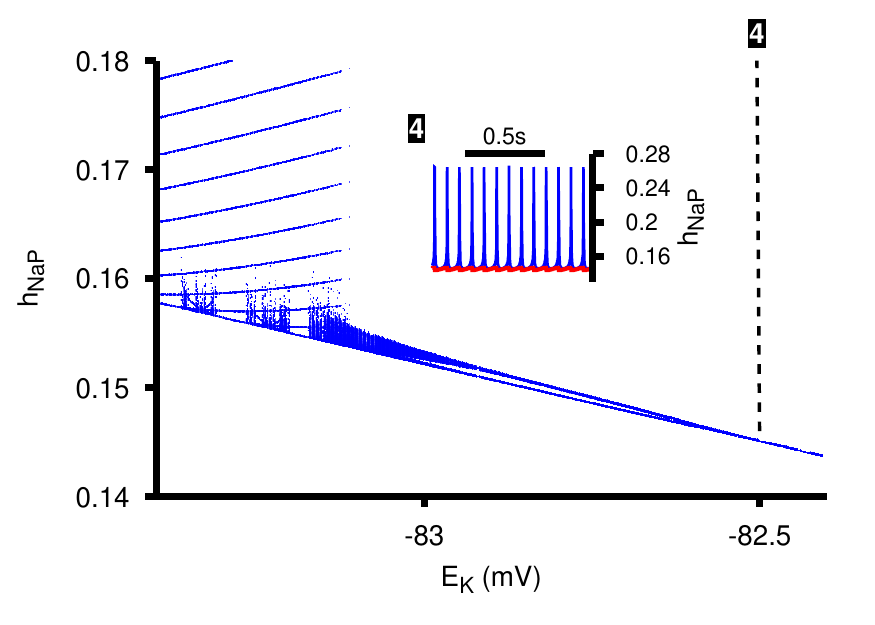}
					\caption*{({\bf C})}
				\end{minipage} 
				\caption{\small{Bifurcation diagram of attracting dynamics of the neuronal model with $[K ^{+}]_{out}$ (and hence $E_{K}$) as the bifurcation parameter, varied in steps of $0.001$ \si{\milli\volt}. (A) Bifurcation over the entire bursting interval. Insets show voltage and $h_{NaP}$ time courses at the fixed values of $E_K$ marked by the numbered vertical dashed lines on the diagram. (B-C) Zoomed views of different parts of the diagram in (A).}}
				\label{chaos}
			\end{figure*}

			Before we move on to incorporate the dynamics of $E_K$ back into the picture, we construct a bifurcation diagram to present in more detail the changes in stable periodic behavior that occur with $E_{K}$ as a bifurcation parameter. More specifically, when the neuronal system exhibits bursting, each burst is composed of a finite number of action potentials, each associated with an approximately constant $h_{NaP}$.  For each fixed $E_K$, for the corresponding periodic spiking or bursting attractor, we therefore record the $h_{NaP}$ value at which each spike occurs (Fig. \ref{chaos}).
			
			In this bifurcation diagram, each dot denotes the value of $h_{NaP}$ at which an action potential occurs during $80$ seconds of simulated bursting behavior, for a corresponding fixed value of $E_K$.  For each $E_K$, the spikes from the first $55$ seconds of neuron simulation are not shown, such that the diagram omits the transient state and only reflects the attractors of the system.  For sufficiently low $E_K$, the stable dynamics consists of periodic tonic spiking, characterized by a single $h_{NaP}$ value for each $E_K$ in the diagram. As $E_K$ increases, the transition from a tonic spiking state to a bursting state appears to arise through a chaotic period doubling mechanism (Fig. \ref{chaos}A,B), estimated numerically to occur just above $E_K = -91.2$ \si{\milli\volt}.
			
			The transition from bursting back to tonic spiking, depicted in Fig. \ref{chaos}A,C, is less clear cut. The spike branch at highest $h_{NaP}$ values seems to disappear instantly as $E_K$ increases. We expect that this change is related to the phenomena shown in Figs. \ref{behavior}C, \ref{2dphase}D.  In the solution displayed in Fig. \ref{behavior}C, it appears that bursting is about to begin, but instead a plateau of  depolarization block occurs.  From Fig. \ref{2dphase}D, we can appreciate that the AH point has moved to smaller $h_{NaP}$ than that of the fold point, such that the trajectory's initial jump to the active phase does not yield a full spike.  Only after $h_{NaP}$ drifts to lower values, below the AH point, can spiking ensue. With an additional increase in $E_K$ to just below $-83.1$ \si{\milli\volt}, most of the remaining spike branches disappear together, leaving only a cluster of values near $h_{NaP}=0.155$.  We also notice pockets of variability in $h_{NaP}$ as $E_K$ varies between $-83.4$ and $-83.1$ \si{\milli\volt}. Interestingly, inspection of the voltage trace suggests that periodic spiking begins at about $E_K = -82.6$ \si{\milli\volt}, above the value at which most of the collection of $h_{NaP}$ branches disappears.
			
			Elucidating the details of this bifurcation is beyond the scope of our consideration of ramping bursts in the  full model and remains for future inquiry, which would require more detailed simulations and analysis.
			
		\subsection{Dynamics in $(E_K, V, h_{NaP})$ phase space}
\begin{figure*}[htbp] 
				\centering
				\begin{minipage}[b]{0.5\linewidth}
					\centering
					\includegraphics[width=1\linewidth,height=1.2\linewidth]{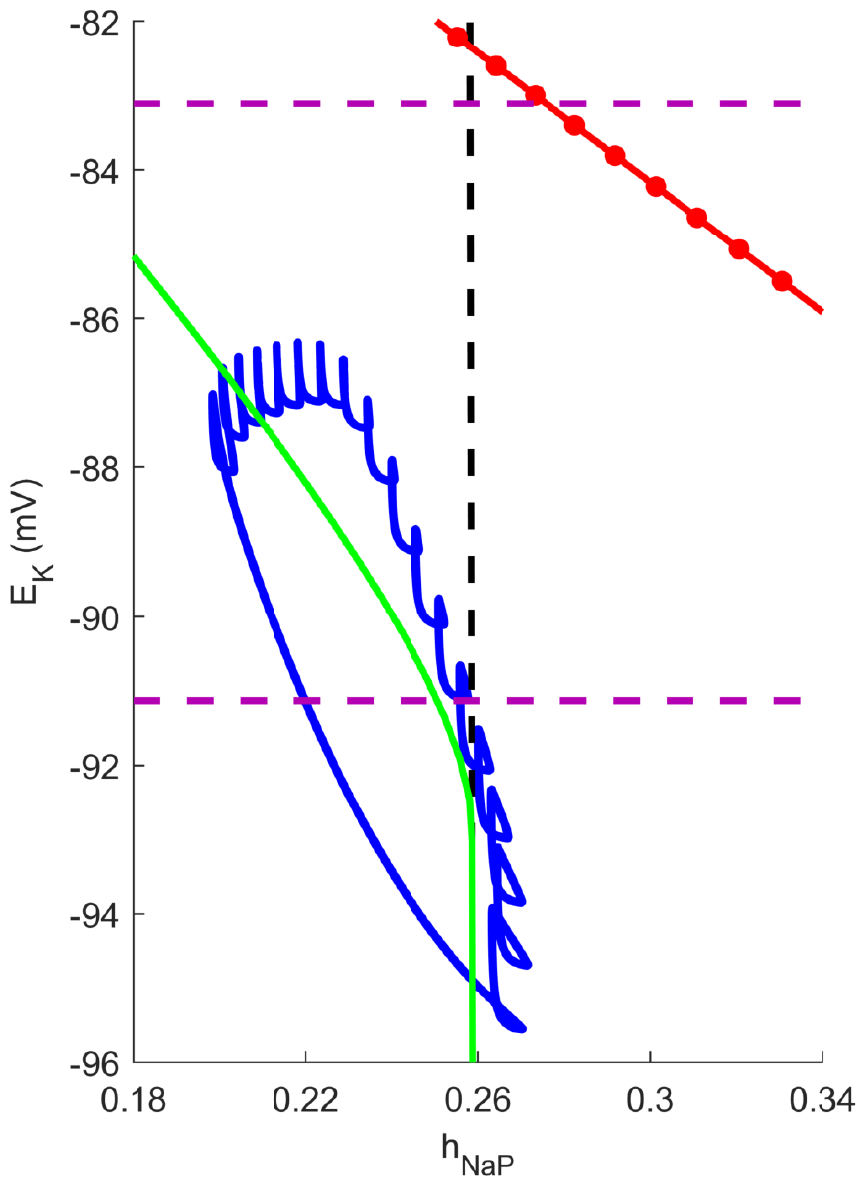} 
					\caption*{({\bf A}): $g_{NaP} = 5.0$ \si{\nano\siemens}, $g_{L} = 2.5$ \si{\nano\siemens}, $g_{syn} = 0.365$ \si{\nano\siemens}}
				\end{minipage}
				\begin{minipage}[b]{0.5\linewidth}
					\centering
					\includegraphics[width=1\linewidth,height=1.2\linewidth]{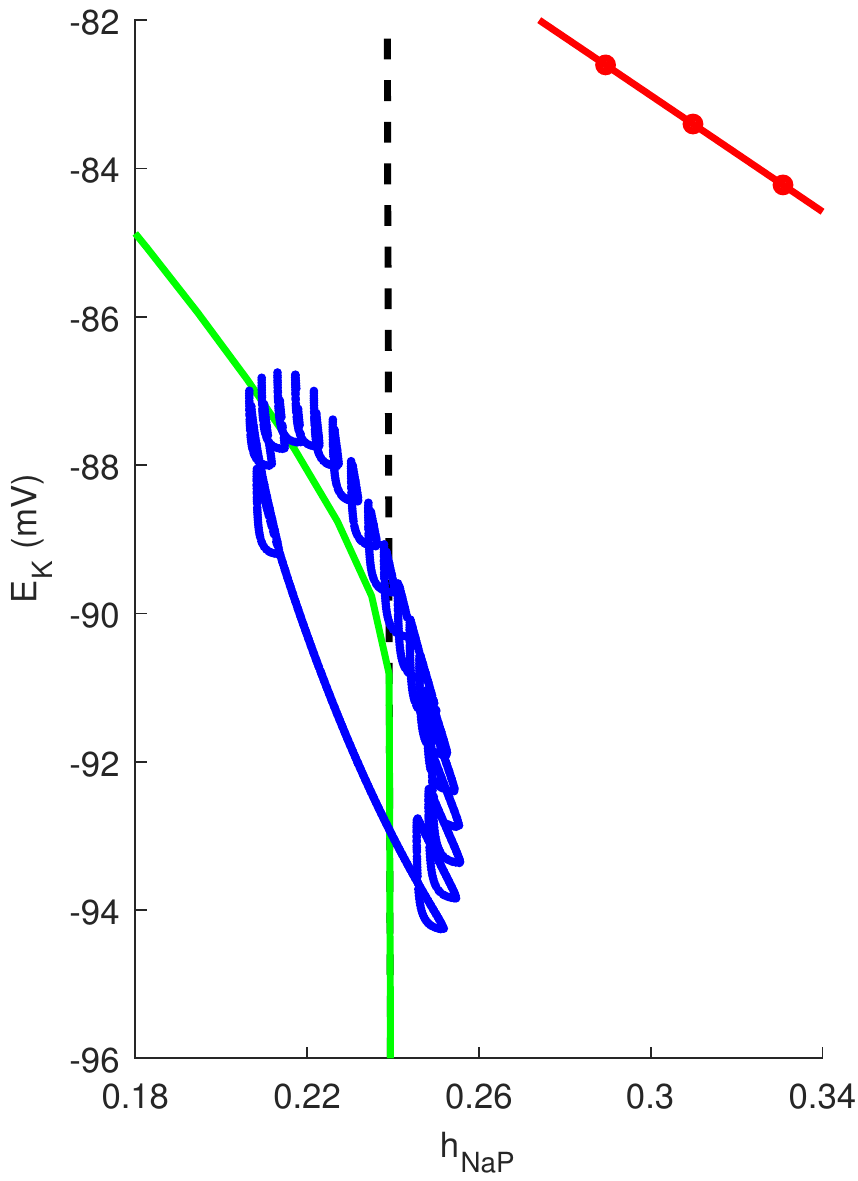}
					\caption*{({\bf B}): $g_{NaP} = 4.5$ \si{\nano\siemens}, $g_{L} = 2.4$ \si{\nano\siemens}, $g_{syn} = 0.360$ \si{\nano\siemens}}
				\end{minipage}
				\caption{\small{Projection of the full system bursting trajectory (blue) to the $\left( E_{K}, h_{NaP} \right)$ plane.  This plane is useful for visualizing the curve of fast subsystem fold points, which is independent of $E_K$ (black dashed line); the curve of fast subsystem AH points (solid-dotted red), which do not play a strong role in the bursting pattern; the fast subsystem periodic orbit termination curve (solid green), which switches from a SNIC, where is aligns with the fold line, to a homoclinic, where it deviates from the fold; and the values of $E_K$ where the neuronal dynamics, with $E_K$ frozen, transitions from spiking to bursting (lower purple dashed line) and from bursting back to spiking (upper purple dashed line). In \textbf{(A)}, we examine the default parameter set, the bursting behavior of which is depicted in Fig. \ref{bursting}(A). In \textbf{(B)} we depict the same trajectory for the parameter set in Fig. \ref{bursting}(C). Note that this parameter set exhibits ramping bursts, despite not showing bursting behavior for any fixed value of $[K^{+}]_{out}$. Thus, there are no dotted-purple lines, as there are no transitions in and out of a bursting state for fixed $[K^{+}]_{out}$.}}
				\label{phase2b}
			\end{figure*}

\begin{figure*}[htbp]
				\begin{minipage}[b]{0.5\linewidth}
					\centering
					\includegraphics[width=1\linewidth]{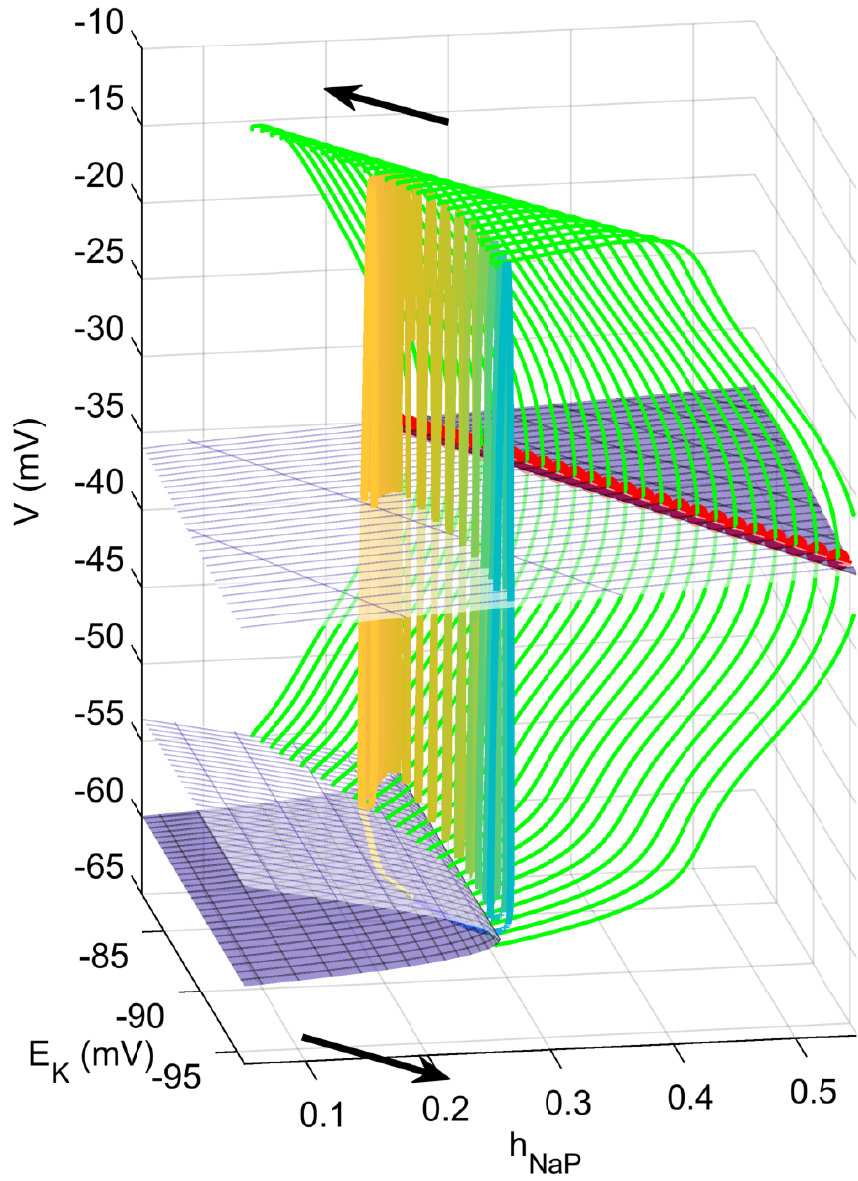}
					\caption*{({\bf A})}
				\end{minipage}
				\begin{minipage}[b]{0.5\linewidth}
					\centering
					\includegraphics[width=1\linewidth]{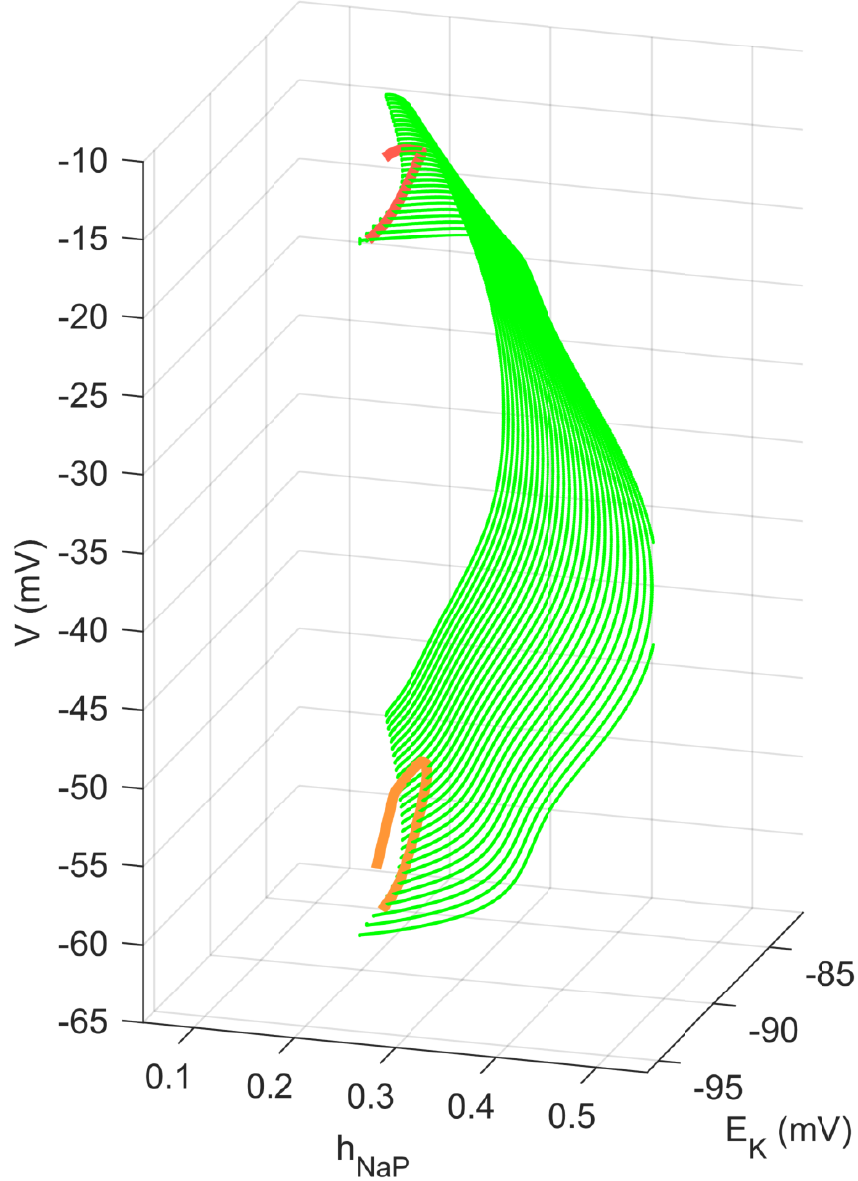}
					\caption*{({\bf B})}
				\end{minipage}
				\caption{\small{The trajectory of a bursting neuron in $\left( E_{K}, V, h_{NaP} \right)$ phase space. \textbf{(A)} Phase space features shown include fast subsystem equilibria (upper and lower blue surfaces), AH curve (red), and maximum and minimum voltages along periodic orbit families emerging from the AH curve (green). The full  model trajectory is also shown projected to this space, color coded from yellow (less negative $E_K$) to blue (more negative $E_K$).  The arrows show the direction of evolution of the bursting trajectory. \textbf{(B)} By tracing the minimum (orange) and maximum (red) values of voltage attained in every spike within the burst, it becomes clear that the trajectory of the neuron travels along the family of periodic orbits during the burst and experiences a decline in spike amplitude when it pulls away from the edge of the periodic orbit family where it starts and terminates (endpoints of the orange curve with larger and smaller $h_{NaP}$, respectively).}}
				\label{phase3d}
			\end{figure*}

			Next, we incorporate the dynamics of $E_K$ back into the picture.  
			Consider the trajectory of the full model system projected into the $(h_{NaP}, E_K)$ plane (Fig. \ref{phase2b}, blue curve), where it progresses in a counterclockwise fashion.\label{key}
			Starting from the quiescent state (the leftmost intersection of the blue neuronal trajectory and the lower purple line), $h_{NaP}$ increases until the trajectory crosses the lower fold  of $\mathcal{S}$, the fast subsystem critical manifold (Fig. \ref{phase2b}, black line), which also corresponds to a the termination of the fast subsystem periodic orbit family (Fig. \ref{phase2b}, green curve).
			If $E_K$ were frozen, then this crossing would result in tonic spiking.  Instead, $E_K$ increases as spiking continues. Eventually $E_K$ crosses the value where the neuronal dynamics supports bursting (Fig. \ref{phase2b}, lower purple line).  Interestingly, we see that very close to this $E_K$, the periodic orbit termination curve diverges from the fold line, confirming that the switch from spiking to bursting in the $E_K$-frozen system corresponds to a switch from termination of the periodic family in a SNIC bifurcation to termination in a homoclinic bifurcation.  As $E_K$ continues to increase, the trajectory moves away from the homoclinic curve and towards the AH curve (Fig. \ref{phase2b}, red line with dots). Oscillation amplitude shrinks to zero at an AH bifurcation.
			Correspondingly, the approach of the trajectory towards the AH curve yields the decrease in spike height seen in Fig. \ref{bursting}A (see also Fig. \ref{phase3d}), representing a less extreme form of the amplitude modulation in the burst patterns arising  with $E_K$ fixed between $-96$ and $-90$ \si{\milli\volt} (Fig. \ref{2dphase}) and in bursting associated with the CAN current in past work \citep{rubin2009,dunmyre,wang2016,wang2020}.  Eventually, $E_K$ peaks and then decays slightly due to $I_{glia}$ and $I_{diff}$, and the decrease in $h_{NaP}$ pulls the trajectory back across the periodic  orbit termination curve, terminating the active phase of the burst.
			
			Putting everything together, we see that the full model system with dynamic $E_K$ engages in a form of parabolic bursting \citep{ermkop1986,rinzel1987}.  Parabolic bursting was originally identified as a form of bursting in which the evolution of two slow variables switches the fast subsystem back and forth across a SNIC curve twice per cycle, yielding an alternation between a quiescent regime corresponding to each inter-burst interval   and a spiking regime corresponding to the active phase of each burst.  This form of bursting was dubbed parabolic in reference to the parabolic shape of the curve depicting spike frequency versus time within each burst, resulting from the low frequency spiking associated with passage near a SNIC bifurcation.  In our case, the use of projection  shows that the initial slow spiking at the start of the burst active phase corresponds to the slow tonic spiking seen with very hyperpolarized $E_K$ (Fig. \ref{behavior}A), which emerges as the trajectory evolves near the fast subsystem SNIC bifurcation curve when $E_K$ is low.  Interestingly, the transition from a SNIC to a homoclinic bifurcation curve  here differs from classical parabolic bursting and accounts for the spike acceleration within the burst and the lack of the significant slowing at the end of the burst typically seen (Fig. \ref{bursting}), consistent with other recent work emphasizing the quantitative variability that can occur within individual bursting classes \citep{rok}.
			Finally, to provide one more perspective that confirms the nature of the bursting dynamics, we visualize the bursting trajectory in the $\left( E_{K}, V, h_{NaP} \right)$ phase space.
			The AH points determined by different $E_{K}$ values can be collected into a curve in $\left( E_{K}, V, h_{NaP} \right)$ space. Similarly, the points of maximum and minimum voltage along the periodic orbit families, parameterized by both $h_{NaP}$ and $E_K$, can be stacked into a surface in this phase space, which is bisected by the AH curve. The trajectory of the full model in this phase space along with these additional structures are depicted in Fig. \ref{phase3d}, which also gives another perspective on the transition from a termination of the fast subsystem periodic solution family in a SNIC to termination in a homoclinic.
			
			To summarize this whole section, our model utilizes persistent sodium currents \citep{butera} and dynamic ion concentrations \citep{barreto} to recreate the ramping preinspiratory / inspiratory behavior seen in bursting \pbc\ neurons. Our model is built from a model proposed in previous work \citep{bacak}, with the addition of dynamic ion concentrations and neuronal regulators \citep{barreto}. The process of bursting in our model can be understood to be a form of parabolic bursting based on  two-dimensional projections, fast-slow decomposition and computation of bifurcation curves, and can be visualized fully by graphing in the $\left( E_{K}, V, h_{NaP} \right)$ phase space.  Ramping of spike frequency at burst onset depends on the passage  of the bursting trajectory near a curve of  SNIC bifurcations that terminates a family of fast subsystem periodic orbits and its subsequent departure from this curve, which prevents a symmetric spike deceleration at the end of each burst.  This burst mechanism does not require there to be a fixed value of $E_K$ at which the remaining equations produce bursting (Figs. \ref{bursting}C-D, \ref{phase2b}B).    The change in spike heights during the burst depends on how the trajectory travels relative to the AH bifurcation curve that gives rise to the periodic orbits.
			
	\section{Robustness of Model Dynamics} \label{section:robustness}
		\subsection{Robustness in conductance parameters}
\begin{figure*}[htbp] 
				\begin{minipage}[b]{0.5\linewidth}
					\centering
					\includegraphics[width=1\linewidth]{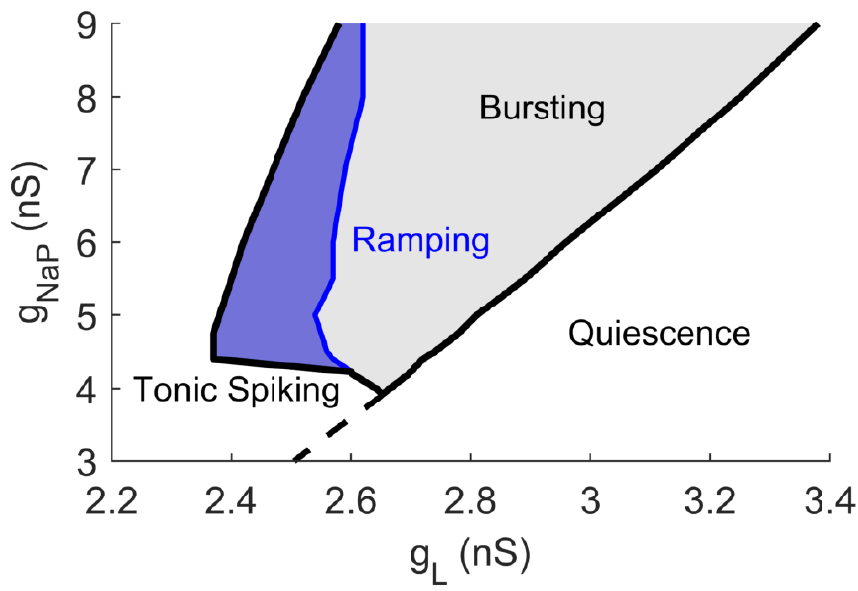} 
					\caption*{({\bf A})} 
				\end{minipage}
				\begin{minipage}[b]{0.5\linewidth}
					\centering
					\includegraphics[width=1\linewidth]{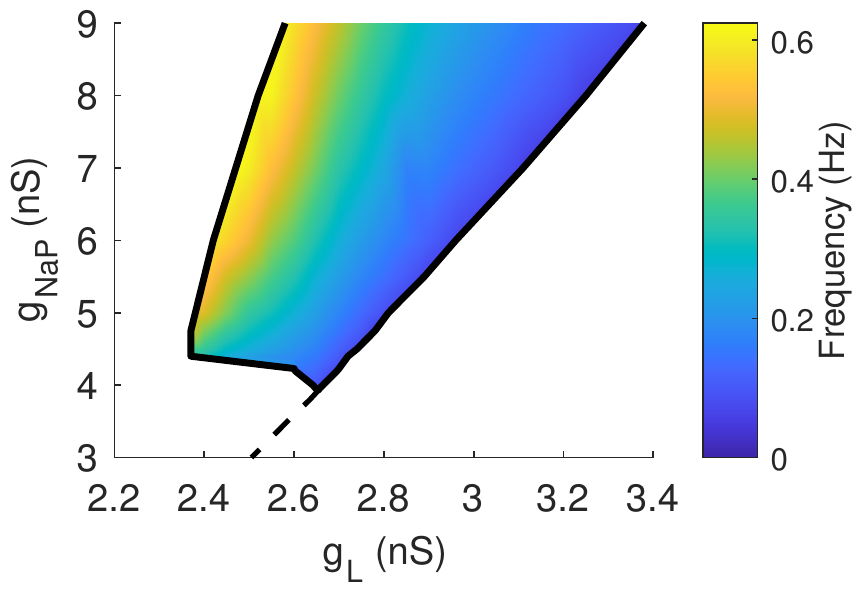}
					\caption*{({\bf B})} 
				\end{minipage}
				\caption{\small{Bursting within the $\left( g_{L}, g_{NaP} \right)$ parameter space. \textbf{(A)} The gray region depicts the set of parameters for which bursting occurs. Within this region, a smaller set of parameters (blue), associated with relatively low $g_{L}$ values, correspond to ramping bursts. For this diagram, a ramping burst was defined as a burst where the external potassium ion concentration $\left( [K^{+}]_{out} \right)$ after the first three spikes is less than the $[K^{+}]_{out}$ required to induce bursting behavior for the model with the same conductances but with a the potassium ion concentration. Parameter sets with lower $g_{L}$ than in the bursting region correspond to tonic spiking behavior, while higher $g_{L}$ led to quiescence. \textbf{(B)} The  burst frequencies for parameter values within the bursting region is indicated by the gradient bar, with more yellow regions corresponding to greater frequencies.}}
				\label{gl_gnap_rob}
			\end{figure*}

			A critical question for any model in which the details of an activity pattern are important is robustness to variation in parameters. Experimental results have confirmed that the presence of persistent sodium $\left( I_{NaP} \right)$ and leakage $\left( I_{L} \right)$ currents are essential to pacemaker activity in \pbc\ neurons \citep{delnegro2002, koizumi2010}. 
			Thus, we mapped the behavior of the model in the $\left( g_{L}, g_{NaP} \right)$ parameter space to measure the robustness of bursting within the neuron under variation of these parameters (Fig. \ref{gl_gnap_rob}A). 
			While bursting behavior could be achieved over a wide range of physiologically relevant parameter values, ramping bursts were restricted to a smaller parameter set.
			Furthermore, we also measured bursting frequency within the bursting parameter region (Fig. \ref{gl_gnap_rob}B), demonstrating how the properties of the model bursting  patterns are modulated by these conductance levels. 
			
\begin{figure*}[htbp] 
				\begin{minipage}[b]{0.5\linewidth}
					\centering
					\includegraphics[width=1\linewidth]{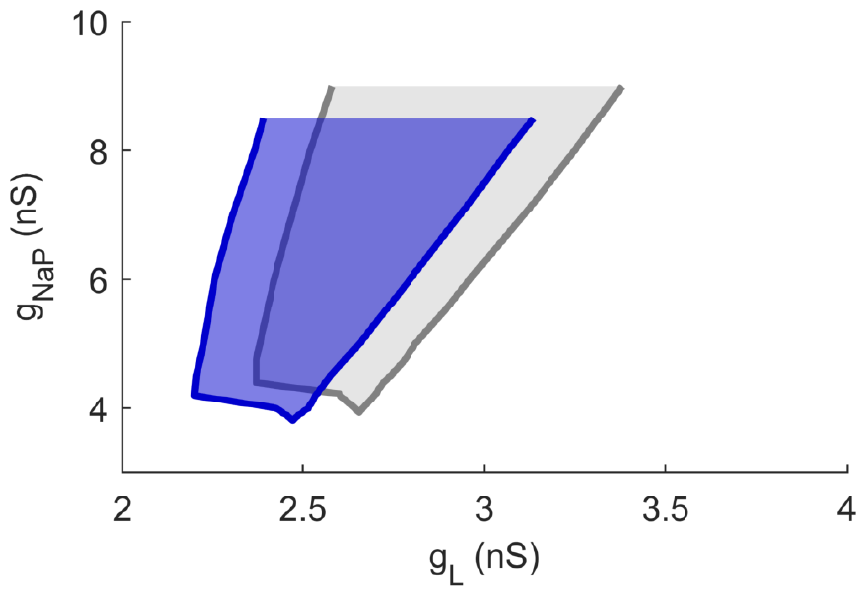}
					\caption*{({\bf A}): $g_{syn} = 0.330$ \si{\nano\siemens}.}
				\end{minipage}
				\begin{minipage}[b]{0.5\linewidth}
					\centering
					\includegraphics[width=1\linewidth]{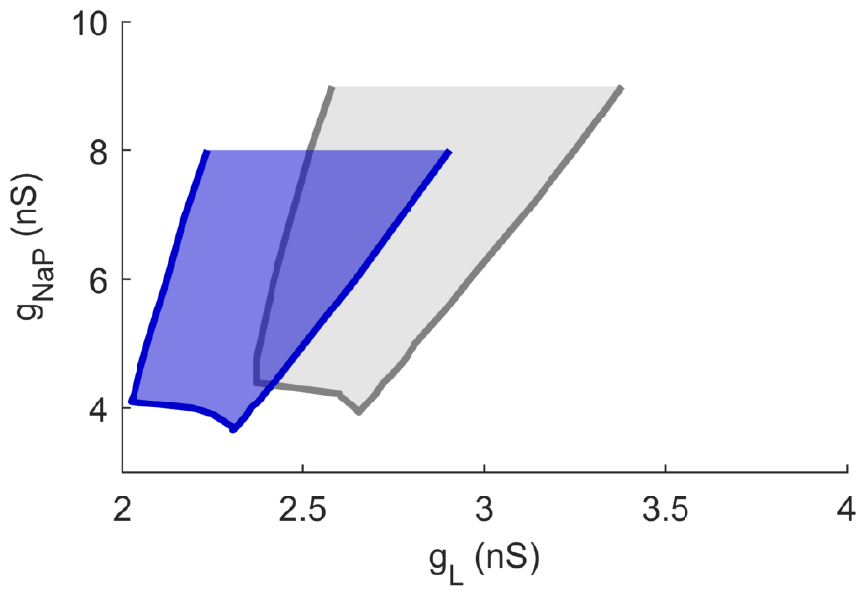}
					\caption*{({\bf B}): $g_{syn} = 0.300$ \si{\nano\siemens}.}
				\end{minipage}
				\begin{minipage}[b]{0.5\linewidth}
					\centering
					\includegraphics[width=1\linewidth]{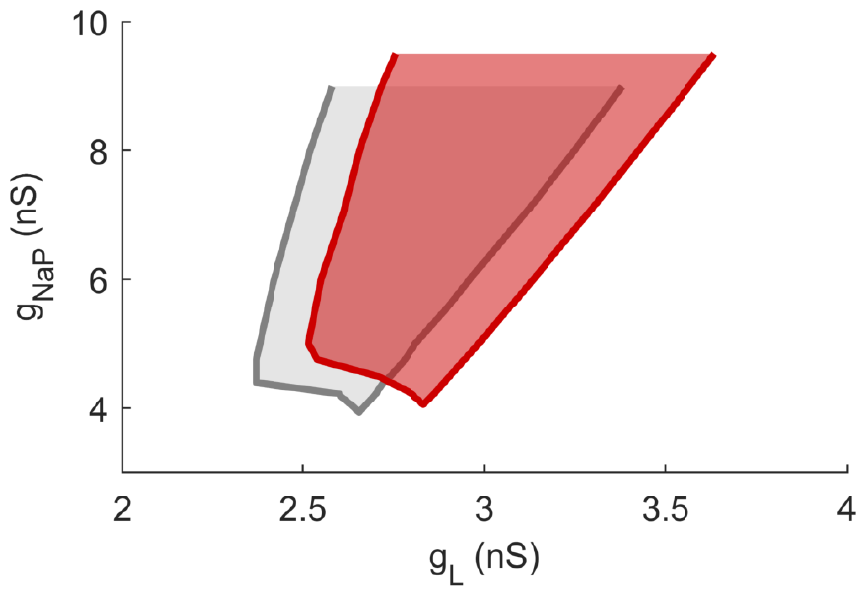}
					\caption*{({\bf C}): $g_{syn} = 0.400$ \si{\nano\siemens}.}
				\end{minipage}
				\begin{minipage}[b]{0.5\linewidth}
					\centering
					\includegraphics[width=1\linewidth]{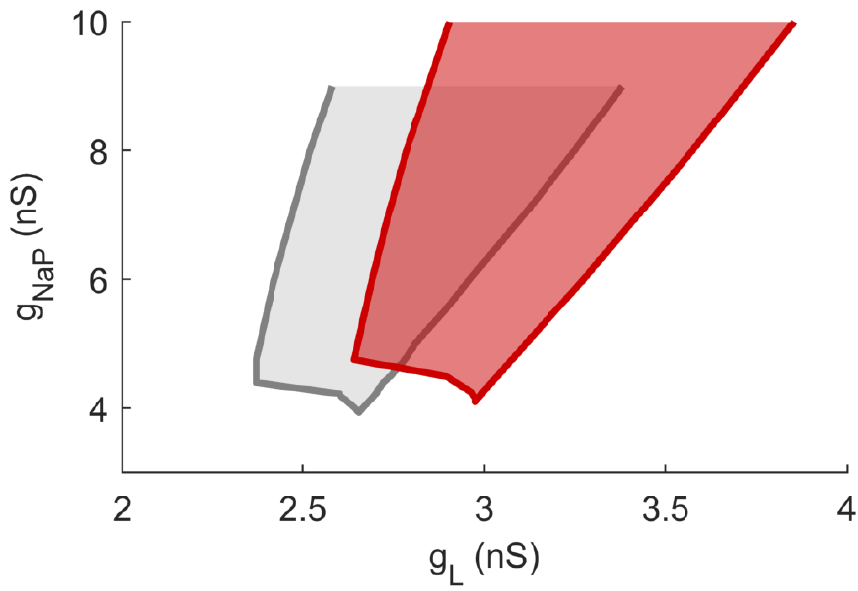} 
					\caption*{({\bf D}): $g_{syn} = 0.430$ \si{\nano\siemens}.}
				\end{minipage}
				\caption{\small{The bursting region within the $\left( g_{L}, g_{NaP} \right)$ parameter space depends on $g_{Syn}$. In all panels, the gray region represents the bursting region for the default parameter value $g_{Syn} = 0.365$ \si{\nano\siemens}. \textbf{(A), (B)} The blue areas represent the bursting regions for decreased synaptic input. Note that as synaptic input is decreased, the shape of the bursting region remains relatively similar, but shifts towards smaller $g_{L}$ and $g_{NaP}$ values. \textbf{(C), (D)} Similarly, the red areas represent the bursting region for increased synaptic input, which shifts bursting towards larger $g_{L}$ and $g_{NaP}$.}}
				\label{gsyn_robust}
			\end{figure*}

			The parameters that induced bursting behavior were also strongly affected by the synaptic input into the neuron. In this model, this tonic input is represented by the current $I_{Syn}$. The effects of altering synaptic input through variation of $g_{Syn}$ on the bursting region within the $\left( g_{L}, g_{NaP} \right)$ parameter space is depicted in Fig. \ref{gsyn_robust}.
			
			The overall shape of these bursting regions is consistent with previous studies \citep{delnegro2002,purvis2007}, which indicate that pacemaker properties are tied to the $g_{NaP}/g_{L}$ ratio. Consistent with this observation, the upper and lower boundaries of the bursting region for our model are approximately linear within the $\left( g_{L}, g_{NaP} \right)$ parameter space. The exact values of $\left( g_{NaP}, g_{L} \right)$ where bursting occurs in this model differ from those presented in \citep{delnegro2002,purvis2007} and include a narrower range of $g_L$ for each fixed $g_{NaP}$. The difference relative to the modeling work \citep{purvis2007} makes sense as that study used the model of $I_{NaP}$-based bursting proposed in \citep{butera}, which incorporates a different membrane capacitance compared to our model.  Moreover, the experiments for which $K_{bath}$ was reported were performed at elevated $K_{bath}$ \citep{delnegro2002}, which would tend to expand the bursting region to larger $g_L$.
			
		\subsection{Inter-model robustness comparison}
			To further analyze the effectiveness of the proposed model, robustness was compared to two existing models of bursting in \pbc\ neurons. Specifically, we examined two facets of robustness: (1) robustness in parameters, i.e., the ability of the model to maintain bursting behavior over a wide range of physiologically observed parameter values, and (2) robustness in behavior modulation, i.e.,  the ability of the model to demonstrate realistic variation in properties of its activity pattern (including bursting frequency, duration, and duty cycle) as parameter values are varied.
			
			First, the proposed model was compared to the model formulated by \citet{bacak}; structurally, the two models differ only in the fact that our model  includes the dynamics of the external potassium ion concentration. Thus, this comparison demonstrates how the introduction of a dynamic ion concentration, which allows for ramping bursts to occur, affects overall robustness. Next, the proposed model was compared to the model introduced in \citet{butera}, which has been incorporated into multiple subsequent computational studies. 
			In the original paper, bursting was induced by increasing $E_{L}$, which increased activation in the neuron. To maintain consistency with the general literature, however, we keep $E_{L}$  fixed and gradually decrease $g_{L}$  to increase activation, and we examine robustness  within the $\left( g_{L}, g_{NaP} \right)$ parameter space. 
			
\begin{figure*}[htbp] 
				\begin{minipage}[b]{0.33\linewidth}
					\centering
					\includegraphics[width=1\linewidth]{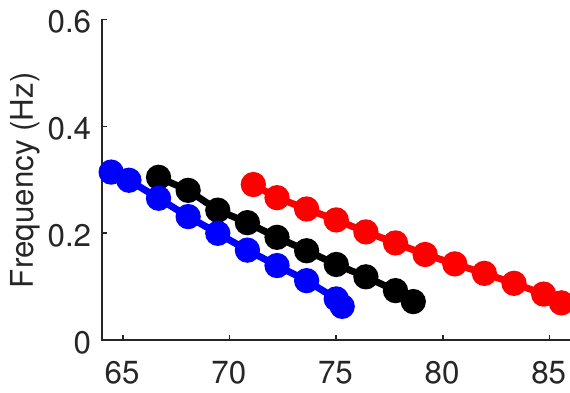}
					\caption*{({\bf A})}
				\end{minipage}
				\begin{minipage}[b]{0.33\linewidth}
					\centering
					\includegraphics[width=1\linewidth]{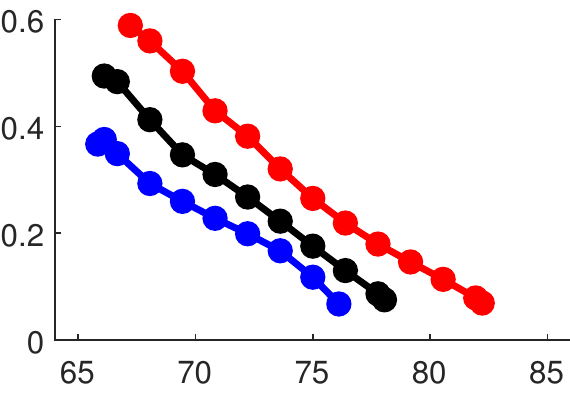}
					\caption*{({\bf B})}
				\end{minipage}
				\begin{minipage}[b]{0.33\linewidth}
					\centering
					\includegraphics[width=1\linewidth]{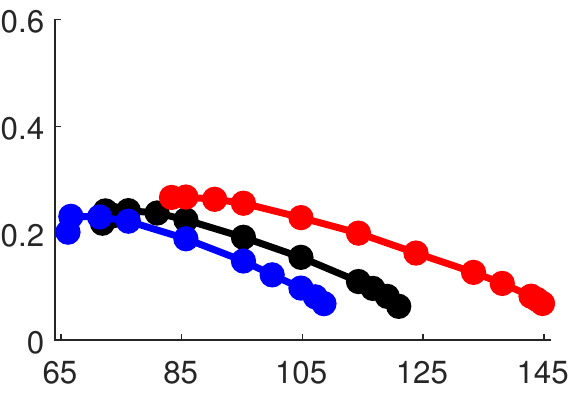}
					\caption*{({\bf C})}
				\end{minipage} \\
				\begin{minipage}[b]{0.33\linewidth}
					\centering
					\includegraphics[width=1\linewidth]{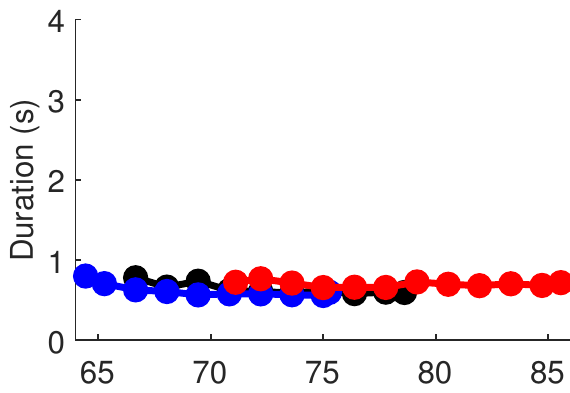}
					\caption*{({\bf D})}
				\end{minipage}
				\begin{minipage}[b]{0.33\linewidth}
					\centering
					\includegraphics[width=1\linewidth]{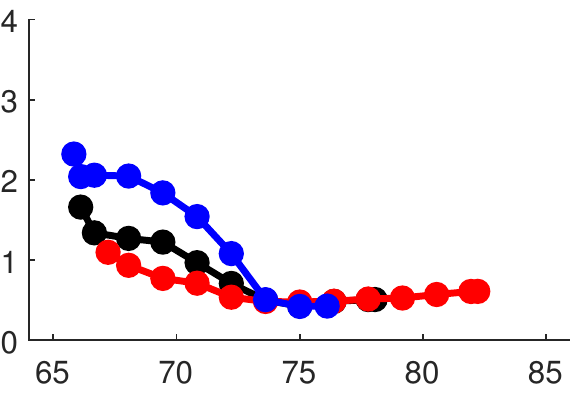}
					\caption*{({\bf E})}
				\end{minipage}
				\begin{minipage}[b]{0.33\linewidth}
					\centering
					\includegraphics[width=1\linewidth]{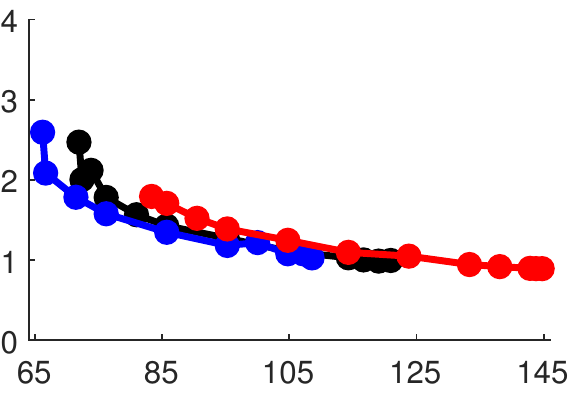}
					\caption*{({\bf F})}
				\end{minipage}\\
				\begin{minipage}[b]{0.33\linewidth}
					\centering
					\includegraphics[width=1\linewidth]{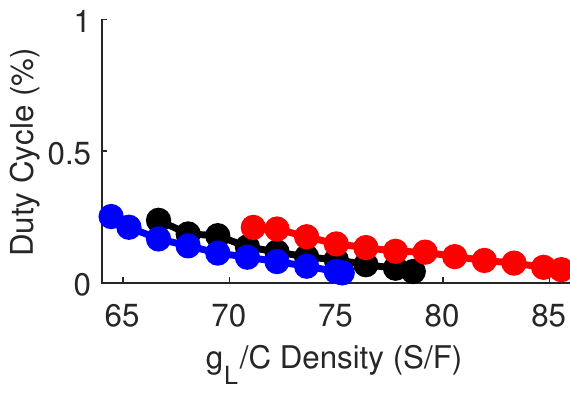}
					\caption*{({\bf G})}
				\end{minipage}
				\begin{minipage}[b]{0.33\linewidth}
					\centering
					\includegraphics[width=1\linewidth]{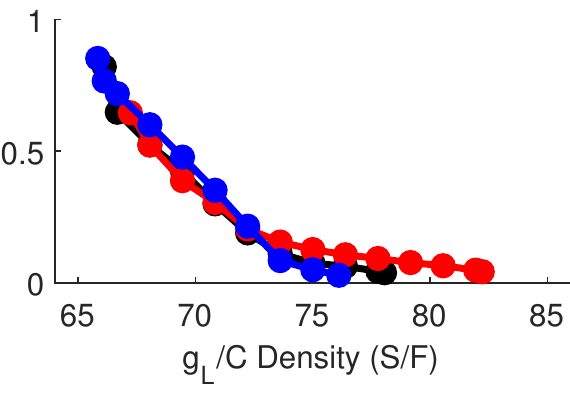}
					\caption*{({\bf H}).}
				\end{minipage}
				\begin{minipage}[b]{0.33\linewidth}
					\centering
					\includegraphics[width=1\linewidth]{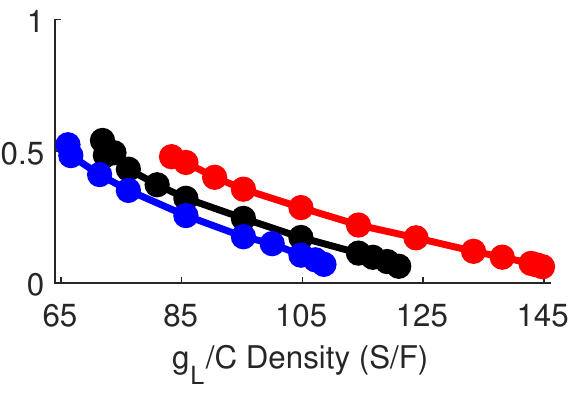}
					\caption*{({\bf I})}
				\end{minipage}
				\caption{\small{The effects of $g_{L}$ and $g_{NaP}$ on the quantitative characteristics of bursting dynamics, compared across models. The left column \textbf{(A, D, G)} represents the model presented in \citet{bacak}, the central column \textbf{(B, E, H)} represents the model proposed in this paper, and the right column \textbf{(C, F, I)} represents the model from \citet{butera}. The top row \textbf{(A, B, C)} shows modulation of bursting frequency, the central row \textbf{(D, E, F)} shows modulation of burst duration, i.e. how long during the burst was the neuron in the spiking state within each burst, and the bottom row \textbf{G, H, I} shows modulation of the neuron's duty cycle, i.e. the ratio of the burst duration to the actual period of the burst. For the proposed model and the model in \citet{bacak}, reduced (blue), default (black), and elevated (red) $g_{NaP}$ values were selected as  $4.5$ \si{\nano\siemens}, $5.0$ \si{\nano\siemens}, and $6.0$ \si{\nano\siemens}, respectively. For the model in \citet{butera}, these conductances were normalized with respect to the reduced membrane capacitance, and set to $2.625$ \si{\nano\siemens},  $2.917$ \si{\nano\siemens}, and $3.500$ \si{\nano\siemens}, respectively. Leakage reversal potential $\left( E_{L} \right)$ was set to $-64$ \si{\milli\volt} for the model in \citet{bacak}, $-68$ \si{\milli\volt} for the proposed model, and $-62$ \si{\milli\volt} for the model in \citet{butera}, based on values in the previous papers and in the earlier parts of this study. Higher bursting frequencies could be reached by fixing $E_{L}$ at a higher value in the model from \citet{butera}, but this strayed even farther from the experimental value of $-68 \pm 3.4$ \si{\milli\volt} determined in \citet{koizumi2010}.}}
				\label{comp}
			\end{figure*}

			The variation of burst properties (frequency, duration, and duty cycle) under changes in $g_{L}$ is depicted for all three models in Fig. \ref{comp}. For each model, this variation was tested for reduced, default, and elevated $g_{NaP}$ values. To adjust for differences between the models, conductance values were normalized with respect to membrane capacitance. Both the proposed model and the model in \citet{bacak} utilize a membrane capacitance of $36$ \si{\pico\farad}, while the model in \citet{butera} utilizes a capacitance of $21$ \si{\pico\farad}. Thus, while the default $g_{NaP}$ value in the proposed model is $5.0$ \si{\nano\siemens}, assuming constant conductance/capacitance density, the default $g_{NaP}$ value in the model in \citet{butera} would be $2.92$ \si{\nano\siemens}.
			
			The first thing to note from this analysis is that compared to the model in \citet{bacak}, the proposed model exhibits bursting behavior over an almost identical set of $g_{L}$ values for the fixed $g_{NaP}$ values tested (Fig. \ref{comp}A,B). Thus, the introduction of a dynamic ion concentration did not alter the robustness of bursting with respect to the $\left( g_{L}, g_{NaP} \right)$ parameter space.
			The inclusion of a dynamic ion concentration significantly increased the set of frequency values attainable through variation of $g_{L}$, however. While the model presented in \citet{bacak} could not reach bursting frequencies above $0.4$ \si{\hertz}, our dynamic potassium model attained bursting frequencies up to $0.6$ \si{\hertz}. It is important to note that, as depicted in Fig. \ref{gl_gnap_rob}, the higher frequency bursts correspond very closely with the newly attainable ramping state.  Moreover, the previous model \citep{bacak} maintained an essentially constant burst duration under variation of both $g_L$ and $g_{NaP}$, whereas our model could achieve substantially longer bursts at the low end of the bursting range of $g_L$ (Fig. \ref{comp}D,E).
			It can be concluded that compared to the static ion concentration model in \citet{bacak}, our proposed model allows for greater modulation of bursting properties through the inclusion of the ramping state, without any significant cost to robustness with respect to conductance parameters.
			
			Compared to the model proposed in \citet{butera}, the proposed model had decreased robustness of bursting with respect to $g_{L}/C$ density, as depicted in Fig. \ref{comp}B,C.
			Despite the decrease in this measure of robustness, the proposed model achieves an increased range of burst frequencies compared to the model proposed in \citet{butera}, as shown in Fig. \ref{comp}B,C. Higher frequency bursts were attainable in the model from \citet{butera}, but only with a shift away from physiologically relevant parameters. Our model also achieved a wider range of burst durations than could be produced by the earlier model \citep{butera} (Fig. \ref{comp}E,F); specifically, our model allowed for shorter bursts at high $g_{NaP}$.
			Consistent with previous experimental results \citep{koizumi2010}, the bursting frequency decreased linearly with an increase in $g_{L}$ for all models.
			
			The quality that stood out most about the proposed model was the significant increase in the range of possible duty cycles when compared to other models, as shown in Fig. \ref{comp}G,H,I. The spiking region could be set to account for an extremely low or extremely high percentage of burst duration, based on variation of $g_{L}$, for all levels of $g_{NaP}$. This flexibility was not possible in the alternative models. Our model produced bursts with similar large duty cycles in the ramping regime, with relatively low $g_L$, with longer, lower frequency bursts for smaller $g_{NaP}$ and shorter, faster bursts for larger $g_{NaP}$.  Neither of the other models could achieve this duty cycle.  Our model produced shorter duty cycles for larger $g_L$, again for all $g_{NaP}$, due to a decrease in burst frequency without much change in burst duration, similar to the other models.
			
	\section{Discussion}
		In this study, we present a model developed from previous conductance-based neuron models that induce bursting behavior dependent on a persistent sodium current \citep{butera,bacak}. Our model replicates the observed frequency ramping behavior of \pbc\ neurons, through the inclusion of external potassium ion $\left( [K^{+}]_{out} \right)$ dynamics. Previous studies have indicated the relationship between fixed levels of $[K^{+}]_{out}$ and burst frequency and duration \citep{delnegro2001}. The incorporation of $[K^{+}]_{out}$ dynamics as an additional slow component of the model induced a modulation of spike frequency throughout the spiking regime, resulting in a robust parabolic bursting behavior.
		
		The dynamics of the model was analyzed through a three-dimensional extension of the traditional fast-slow decomposition. Steady-state behavior was plotted in the $\left(V, h_{NaP} \right)$-phase space for various fixed values of $[K^{+}]_{out}$ and in of turn $E_{K}$, via equation (\ref{eq:nernst}). The curves of saddle-node and AH bifurcations and the families of periodic orbits originating from the AH points were mapped with respect to $E_{K}$. These quantities were then projected onto the $\left( E_{K}, V, h_{NaP} \right)$-phase space, depicting the geometry that ultimately dictates bursting dynamics. Tracking the trajectory of the burst through this phase space revealed that oscillations in the $\left(V, h_{NaP} \right)$-phase space gradually drive the trajectory to higher values of $E_{K}$; subsequently, higher $E_{K}$ values correspond to higher frequency spiking, causing a positive feedback loop resulting in ramping bursts.
		Eventually, the $E_{K}$ level saturates due to the nonlinear dependence of processing of potassium ions by glia, at which point the slow inactivation of $I_{NaP}$ can terminate the burst.  Specifically, as $h_{NaP}$ decays, the fast subsystem periodic orbit family terminates in a homoclinic bifurcation and the voltage repolarizes, corresponding to a transition to the quiescent state of the burst cycle. 
		Finally, the lack of spiking activity causes $E_{K}$ to decay back to a baseline level as the trajectory of the neuron travels back to the saddle-node bifurcation curve, where it re-enters the spiking regime of the burst.  Hence, the dynamical system yields  parabolic bursting behavior that terminates in a homoclinic bifurcation.
		
		In classic parabolic bursting, burst initiation occurs when the trajectory induced by the dynamics of the slow subsystem, which includes two or more slow variables, crosses a SNIC bifurcation curve for the fast subsystem.  As fast spikes ensue, the trajectory of the averaged slow equations eventually progresses back across the SNIC curve, terminating the active phase of the burst.  Thus, the spikes near both burst onset and burst termination are slower than those in the heart of the burst, resulting in a parabolic dependence of spike frequency on spike number within the burst \citep{rinzel1987}.  This paper adds to the collection of past works that have included variations on this structure, including crossings of additional fast subsystem bifurcation curves during the active phase, which result in corresponding variability of burst profiles \citep{rubin2009,barreto,rok}. Specifically, due to the interplay of dynamic $E_K$ and $I_{NaP}$, the ramping bursts in our model terminate via a crossing of a homoclinic bifurcation curve for the fast subsystem, rather than a SNIC, with little slowing of spiking at the end of the burst.  In theory, a homoclinic crossing should also be associated with some spike slowing, but the quantitative details are system-specific \citep{rok}. Future work to extend this model to take into account dynamics of other ion concentrations, in addition to $[K^+]$, may yield even more diverse burst profiles (cf. \citet{barreto}). Specifically, in addition to $Na^+$ dynamics, the dynamics of $Cl^-$ is an often-overlooked factor that could contribute to ramping bursts \citep{currin2020,pace2007}.  Importantly, concentrations of ions that impact neuronal dynamics can be coupled through pumps that transport multiple ion types, so modeling the details of this dynamics in the context of neuronal bursting represents an interesting challenge for future work.
		
		Analysis of model robustness revealed multiple insights. In comparison to previous models of \pbc\ neuron dynamics, the model proposed in this paper exhibits similar robustness with respect to variations in parameters, while offering a greater degree of modulation of burst geometry characteristics, such as frequency, duration, and duty cycle. 
		One exception is that our model's bursting behavior does not extend over the full range of $g_{L}/C$ over which bursting occurs in the  model by \citet{butera}. The robustness of \pbc\  bursting to $g_L/C$ has not been experimentally tested, however, and in fact experiments suggest that not the leak conductance itself but rather the ratio $g_{NaP}/g_L$ is what matters for whether bursting occurs or not \citep{delnegro2001,delnegro2002,purvis2007} (cf. the nearly linear boundaries of the bursting regions in our Figs. \ref{gl_gnap_rob}, \ref{gsyn_robust}).   
		While our decision to treat $E_L$ as a constant allowed us to compare our model directly to earlier ones where leak strength was used to explore model behavior, 
		$E_{L}$ may in reality be nonlinearly modulated by ion  dynamics \citep{koizumi2008,huang2015}. The robustness of bursting that we found with respect to variations in $g_L$ supports the claim that the ramping dynamics that we have studied will persist with the inclusion of $E_L$ dependence on dynamic ion concentrations, but incorporating this effect in the model and tuning it appropriately is beyond the scope of the current study. 
		Another future direction will be the inclusion of additional membrane currents, such as $I_A, I_{KCa}$, the Ca$^{2+}$-activated nonspecific cation (CAN) current, and the $Na/K$ pump current, which have been shown to have a significant effect on \pbc\ neuron and network dynamics in multiple past experimental and computational works \citep{hayes2008,pace2007,zavala2008calcium,krey2010outward,jasinski,koizumi2018transient,picardo2019trpm4,rubin2009,dunmyre,phillips2018dendritic,phillips2019}.
		
		The results of this study reveal a potential role of dynamic ion concentrations in producing and shaping ramping behavior within neuronal bursting. Previous computational studies of \pbc\ neuron activity have ignored the dynamics of $[K^{+}]_{out}$, modeling it as a fixed parameter. This viewpoint has been utilized in experimental studies as well, where $[K^{+}]_{out}$ has often been viewed as equivalent to the potassium concentration of the solution used to bathe slices of neural tissue during experimentation ($k_{bath}$). Our study implies that the physiologically observed oscillations of $[K^{+}]_{out}$ can have a significant impact on \pbc\ neuron dynamics; moreover, similar effects could emerge in prolonged bursting behavior of other neurons and should be incorporated in corresponding models in future work. 
		Rather than assuming that $k_{bath}$ = $[K^{+}]_{out}$, our model incorporates $k_{bath}$ as an environmental factor that can affect the dynamics of $[K^{+}]_{out}$ via diffusion, following the framework of previous computational models that considered dynamic ion concentrations \citep{barreto}. The role of  $k_{bath}$ is important because $k_{bath}$ can be modulated experimentally, providing a way for the mechanism proposed in this paper to be tested. If $[K^{+}]_{out}$ governs ramping dynamics through the mechanism proposed in this paper, then it would be expected that increasing $k_{bath}$ would cause a much faster build-up of $[K^{+}]_{out}$, translating to a faster increase in spiking frequency throughout the burst and hence a steeper frequency ramp. Adjusting $k_{bath}$ to lie above some level would remove the ramping effect all together, as $[K^{+}]_{out}$ buildup would primarily be driven by the excess influx of external potassium from the bathing solution, rather than the export of internal potassium during spiking. Similarly, lowering $k_{bath}$ should correspond to slowing the rise in frequency over the course of the burst. Setting $k_{bath}$ below some threshold  would cause the rate of removal of $[K^{+}]_{out}$ via diffusion to increase enough to entirely prevent the $[K^{+}]_{out}$ buildup needed to induce a bursting state.  It is our hope that future experiments consider the effects of potassium concentration in the bathing solution on the dynamic ramping behavior of individual neurons, to test the mechanisms proposed in this paper. A complication, however, is that prolonged changes in $k_{bath}$ may induce other compensatory effects \citep{okada2005,ransdell2012,he2020}.
		
		Another important future research direction related to this work should involve an expansion of the scope of the model, specifically to analyze the effects of ion-dependent ramping on the neuronal control of respiratory rhythms. The intrinsic dynamic mechanisms within individual neurons and synaptic network interactions work together to generate and modulate breathing rhythms (e.g., \citet{molkov2017,delnegro2018,rubin2019,phillips2019effects,phillips2019}). A specific step to link these factors would be to construct a computational network of both pacemaker and non-pacemaker neurons in the \pbc\ and to model the effects of ramping behavior in pacemakers on the recruitment of non-pacemakers, to advance our understanding of the generation and patterning of inspiratory neural bursts \citep{kam2013,kallurkar2020}. 
		A potential approach to the network modeling problem would be to address the inherent limitations of using a system of ODEs in depicting neuronal behavior. Spatial interactions, which can be an important factor in network dynamics, are not captured by ODE models. This is especially relevant for our model as it incorporates diffusion, a naturally spatially-dependent process, to differentiate between the equilibrium concentration of external potassium ($k_{bath}$) and localized concentration of potassium near the neuronal membrane ($[K^{+}]_{out}$). Hence, one possible research direction would be the development of a network-based model that utilizes both ODEs and PDEs to depict the spiking behavior of individual neurons and spatially dependent processes governing ion dynamics, respectively. 
		The development of the PDE component of the model would have to incorporate $[K^{+}]_{out}$ as both a space- and time-dependent variable, which allows the spiking behavior of each neuron to affect the localized external ion concentrations of its neighboring neuron and is subject to the boundary conditions imposed by the presence of the bathing solution (e.g., a Dirichlet boundary condition forcing $[K^{+}]_{out}$ to take a value of $k_{bath}$ at the boundaries of a modeled brain slice).
		
		To summarize, this paper presents a new model of neuronal bursting in pacemaker neurons, which results in a frequency ramp at bursting onset. This effect was demonstrated to be a manifestation of  parabolic bursting dynamics that allows for a broad range of burst frequencies and duty cycles. The results of this study imply that oscillations in external potassium concentration can play a significant role in the ramping dynamics of \pbc\ neurons. This ion-dependent ramping mechanism should be tested in future experimental studies and incorporated in future models of networks of \pbc\ neurons, and is likely relevant to prolonged bursting dynamics in other neurons and neuronal populations.
		
	\begin{acknowledgements}
		The authors would like to acknowledge the Program in Neural Computation at the Center for the Neural Basis of Cognition for their help in facilitating this research collaboration.
	\end{acknowledgements}

	\clearpage
	\appendix
	\onecolumn
	\normalsize{
		\section{Constants and Parameters} \label{constants}
			The complete list of parameters used for this model is shown below. Certain parameters were fixed for all simulations, while others were varied for different tests. These instances will be noted.\\
			
			Universal \& Experimental Constants:
			\begin{itemize}
				\item Elementary Charge: $q=1.602 \times 10^{-19}$ $C$.
				\item Avogadro Constant: $N_{A} = 6.022 \times 10^{23}$ $\frac{1}{mol}$.
				\item Unit Time Constant: $\tau = 1000$ $\frac{ms}{s}$.
				\item Ratio of Volumes: $\beta = 14.555$ (modified from \citet{barreto}, $\beta = 7$).
				\item Membrane Capacitance: $C = 36$ \si{\pico\farad} (taken from \citet{rybak2007}).
			\end{itemize}
		
			Derived Constants:
			\begin{itemize}
				\item Current Conversion Constant: $\gamma = 7.214 \times 10^{-3}$ $\frac{mM}{s \cdot pA}$ (derived in Appendix \ref{gamma}).
			\end{itemize}
		
			Conductances:
			\begin{itemize}
				\item $\bar{g}_{Na} = 150$ \si{\nano\siemens} (taken from \citet{jasinski}).
				
				\item $\bar{g}_{NaP} = 5$ \si{\nano\siemens} (taken from \citet{bacak}). Varied as parameter in Section \ref{section:robustness}.
				
				\item $\bar{g}_{K} = 160$ \si{\nano\siemens} (taken from \citet{jasinski}).
				
				\item $\bar{g}_{L} = 2.5$ \si{\nano\siemens} (taken from \citet{jasinski}, $\bar{g}_{L} \in [ 2,3 ]$). Varied as parameter in Sect. \ref{section:robustness}.
				
				\item $\bar{g}_{Syn} = 0.365$ \si{\nano\siemens}. (Introduced in this paper to represent constant synaptic drive, in contrast to model in \citet{bacak} where $\bar{g}_{Syn} = 0$). Varied as parameter in Sect. \ref{section:robustness}.
			\end{itemize}
		
			Ion Concentrations \& Reversal Potentials:
			\begin{itemize}
				\item $[Na^{+}]_{out}=120$ \si{\milli\Molar} (taken from \citet{jasinski}).
				
				\item $[Na^{+}]_{in}=15$ \si{\milli\Molar} (taken from \citet{izhikevich}, $[Na^{+}]_{in} \in [5, 15]$).
				\item $E_{Na} = 26.7 \cdot \log{\frac{[Na^{+}]_{out}}{[Na^{+}]_{in}}} = 55.5$ \si{\milli\volt}. (Consistent with \citet{rybak2007}, $E_{Na} = 55$).
				
				\item $[K^{+}]_{in} = 160$ \si{\nano\siemens} (modified from \citet{izhikevich, jasinski}, $[K^{+}]_{in} = 140$).
				
				\item $E_{L} = -68$ \si{\milli\volt} (taken from \citet{jasinski}).
				\item $E_{Syn} = -10$ \si{\milli\volt} (taken from \citet{jasinski}).
			\end{itemize}
		
			Parameters for Fast Sodium $(I_{Na})$ and Persistent Sodium ($I_{NaP}$):
			\begin{itemize}
				\item $V_{m_{Na}} = -43.8$ \si{\milli\volt}, $k_{m_{Na}} = 6$ \si{\milli\volt}, $V_{\tau_{m_{Na}}} = -43.8$ \si{\milli\volt}, $k_{\tau_{m_{Na}}} = 14$ \si{\milli\volt}.
				\item $V_{h_{Na}} = -67.5$ \si{\milli\volt}, $k_{h_{Na}} = -11.8 mV$, $V_{\tau_{h_{Na}}} = -67.5$ \si{\milli\volt}, $k_{\tau_{h_{Na}}} = -12.8$ \si{\milli\volt}.
				\item $V_{m_{NaP}} = -47.1$ \si{\milli\volt}, $k_{m_{NaP}} = 3.1$ \si{\milli\volt}, $V_{\tau_{m_{NaP}}} = -47.1$ \si{\milli\volt}, $k_{\tau_{m_{NaP}}} = 6.2$ \si{\milli\volt}.
				\item $V_{h_{NaP}} = -60$ \si{\milli\volt}, $k_{h_{NaP}} = -9$ \si{\milli\volt}, $V_{\tau_{h_{NaP}}} = -60$ \si{\milli\volt}, $k_{\tau_{h_{NaP}}} = 9$ \si{\milli\volt}.
				
				\item $\bar{\tau}_{m_{Na}} =0.25$ \si{\milli\second}, $\bar{\tau}_{h_{Na}} = 8.46$ \si{\milli\second}, $\bar{\tau}_{m_{NaP}} =1$ \si{\milli\second}, $\bar{\tau}_{h_{NaP}} = 5000$ \si{\milli\second}.
				\item All parameters were taken directly from \citet{bacak2}, with the exception of $k_{h_{Na}}$, which was altered from a value of $-10.8$ \si{\milli\volt} to the listed value of $-11.8$ \si{\milli\volt}.
			\end{itemize}
		
			Parameters for Delayed Rectifier Potassium Current $(I_{K})$:
			\begin{itemize}
				\item $n_{A} = 0.01$ $\frac{1}{mV}$, $n_{A_{V}} = 44$ \si{\milli\volt}, $n_{A_{k}} = 5$ \si{\milli\volt}, $n_{B} = 0.17$, $n_{B_{V}} = 49$ \si{\milli\volt}, $n_{B_{k}} = 40$ \si{\milli\volt}.
				\item All values taken from \citet{bacak}.
			\end{itemize}
			
			Parameters for Diffusion of Extracellular Potassium $([K^{+}]_{out})$:
			\begin{itemize}
				\item $k_{bath}=4$ \si{\milli\Molar} (taken from \citet{barreto}).
				\item $\tau_{diff} = 750$ \si{\milli\second} (numerically equivalent to the formulation in \citet{barreto}, which uses $\frac{1}{\tau_{diff}} \equiv \frac{\epsilon}{\tau}$, where $\epsilon=1.333$ \si{\hertz} and $\tau=1000 \frac{\si{\milli\second}}{\si{\second}}$).
			\end{itemize}
			
			Parameters for Glia:
			\begin{itemize}
				\item $\bar{G} = 10 \frac{\si{\milli\Molar}}{\si{\second}}$, $\bar{K} = 5 \si{\milli\Molar}$, ${z_{K}} = 6 \frac{1}{\si{\milli\Molar}}$.
				\item These parameter values were altered from those in \citet{barreto}. In \citet{barreto}, the concentration of $[K^{+}]_{out}$ remains far below the mid-point value of the sigmoidal equation (\ref{eq:glia}). The parameters were adjusted such that the range of dynamic $[K^{+}]_{out}$ was distributed over the midpoint of equation (\ref{eq:glia}), ensuring that the nonlinear behavior of glial cells was represented.
			\end{itemize}
	
	\clearpage	
	\onecolumn
		\section{Derivation of $\gamma$} \label{gamma}
			Our initial assumption is that the neuron is roughly spherical, or rather that the majority of the cell's volume is contained in a sphere. From \cite{barreto}, the radius of the neuron is taken to be approximately $r=7.0$ \si{\um}. Hence, the internal volume of the neuron can be approximated as $V_{in} = \frac{4}{3} \pi r^3 = 1.44 \times 10^{-9}$ \si{\mL}.
			
			The internal concentration $c_{in}$ can be determined from the total number of total ions $N$, and the internal volume $V_{in}$, and Avogadro's Constant $N_{A}$.
			
			$$c_{in} = N \cdot \frac{1}{N_A} \cdot \frac{1}{V_{in}}$$
			
			Note that the ions we are measuring concentrations of are $Na^{+}$ and $K^{+}$, both of which have an $+1$ charge. Letting $q=1.60 \times 10^{-19}$ \si{\coulomb}, we can express the concentration in terms of total charge, $Q$.
			
			$$c_{in} = \frac{N}{V_{in} N_{A}} \cdot \frac{q}{q} = \frac{Q}{q V_{in} N_{A}}$$
			
			Differentiating, we get:
			
			$$\frac{dc_{in}}{dt} = \frac{d}{dt} \left( \frac{Q}{q V_{in} N_{A}} \right) = \frac{dQ}{dt} \cdot \frac{1}{q V_{in} N_{A}} = I \cdot \frac{1}{q V_{in} N_{A}}$$
			
			By , we can determine:
			
			$$\gamma \equiv \frac{1}{q V_{in} N_{A}} = 7.2 \times 10^{3} \frac{\si{\mole}}{\si{\coulomb} \cdot \si{\mL}}$$
			
			By the following dimensional analysis manipulation, we obtain:
			
			$$\frac{\si{\mole}}{\si{\coulomb} \cdot \si{\milli\liter}} \cdot \left( 10^{3} \frac{\si{\milli\mole}}{\si{\mole}} \cdot 10^{3} \frac{\si{\mL}}{\si{\liter}} \cdot \frac{\si{\second}}{\si{\second}} \cdot \frac{\si{\milli\Molar} \cdot \si{\liter}}{\si{\milli\mole}} \cdot \frac{\si{\coulomb}}{\si{\ampere} \cdot \si{\second}} \cdot 10^{-12}\frac{\si{\ampere}}{\si{\pico\ampere}} \right) = 10^{-6} \frac{\si{\milli\Molar}}{\si{\second} \cdot \si{\pico\ampere}}$$
			
			Thus, we conclude:
			\begin{equation} \label{eq:gamma}
				\gamma = 7.214 \times 10^{-3} \text{ }\frac{\si{\milli\Molar}}{\si{\second}} \cdot \frac{1}{\si{\pico\ampere}}
			\end{equation}
	}

\end{document}